\documentclass[twocolumn]{aastex631}
\usepackage[utf8]{inputenc}
\usepackage[figure,figure*]{hypcap}
\usepackage{multirow}
\usepackage{natbib}
\usepackage{xcolor}
\usepackage{soul}
\usepackage{CJK}

\setstcolor{red}
\hypersetup{linkcolor=red,citecolor=cyan,filecolor=magenta,urlcolor=blue}

\newcommand{\angstrom}{\mbox{\normalfont\AA}}

\newcommand{\ergskpc}{erg~s$^{-1}$~kpc$^{-2}$}
\newcommand{\ergsAkpc}{erg~s$^{-1}$~\AA$^{-1}$~kpc$^{-2}$}

\newcommand{\uvha}{UV-to-H$\alpha$}
\newcommand{\ha}{H$\alpha$}

\newcommand{\hb}{H$\beta$}
\newcommand{\hii}{H~{\small II}}
\newcommand{\oiii}{[O~{\small III}]}
\newcommand{\nii}{[N~{\small II}]}

\newcommand{\swarp}{{\sc SWarp}}
\newcommand{\pypher}{{\sc PyPHER}}
\newcommand{\grizli}{{\sc Grizli}}
\newcommand{\dbasis}{{\sc Dense Basis}}

\shorttitle{Spatially resolved analysis of burstiness by comparing rest-UV and \ha\ at $z\sim1$}
\shortauthors{Mehta et al.}

\begin{document}

\title{A spatially resolved analysis of star-formation burstiness by comparing UV and \ha\ in galaxies at $z\sim1$ with UVCANDELS}

\author[0000-0001-7166-6035]{Vihang Mehta}
\affiliation{IPAC, Mail Code 314-6, California Institute of Technology, 1200 E. California Blvd., Pasadena, CA, 91125, USA}
\email{vmehta@ipac.caltech.edu}


\author[0000-0002-7064-5424]{Harry I. Teplitz}
\affiliation{IPAC, Mail Code 314-6, California Institute of Technology, 1200 E. California Blvd., Pasadena, CA, 91125, USA}

\author[0000-0002-9136-8876]{Claudia Scarlata}
\affiliation{Minnesota Institute for Astrophysics, University of Minnesota, 116 Church St SE, Minneapolis, MN 55455, USA}

\author[0000-0002-9373-3865]{Xin Wang}
\affiliation{School of Astronomy and Space Science, University of Chinese Academy of Sciences (UCAS), Beijing 100049, China}
\affiliation{National Astronomical Observatories, Chinese Academy of Sciences, Beijing 100101, China}
\affiliation{IPAC, Mail Code 314-6, California Institute of Technology, 1200 E. California Blvd., Pasadena, CA, 91125, USA}

\author[0000-0002-8630-6435]{Anahita Alavi}
\affiliation{IPAC, Mail Code 314-6, California Institute of Technology, 1200 E. California Blvd., Pasadena, CA, 91125, USA}

\author[0000-0001-6482-3020]{James Colbert}
\affiliation{IPAC, Mail Code 314-6, California Institute of Technology, 1200 E. California Blvd., Pasadena, CA, 91125, USA}

\author[0000-0002-9946-4731]{Marc Rafelski}
\affiliation{Space Telescope Science Institute, Baltimore, MD 21218, USA}


\author[0000-0001-9440-8872]{Norman Grogin}
\affiliation{Space Telescope Science Institute, Baltimore, MD 21218, USA}

\author[0000-0002-6610-2048]{Anton Koekemoer}
\affiliation{Space Telescope Science Institute, Baltimore, MD 21218, USA}

\author[0000-0002-0604-654X]{Laura Prichard}
\affiliation{Space Telescope Science Institute, Baltimore, MD 21218, USA}

\author[0000-0001-8156-6281]{Rogier Windhorst}
\affiliation{School of Earth \& Space Exploration, Arizona State University, Tempe, AZ\,85287-1404, USA}


\author{Justin M. Barber}
\affiliation{Department of Physics \&\ Astronomy, University of California, Riverside, CA 92521, USA}

\author[0000-0003-1949-7638]{Christopher J. Conselice}
\affiliation{Jodrell Bank Centre for Astrophysics, University of Manchester, Oxford Road, Manchester, UK}

\author[0000-0002-7928-416X]{\begin{CJK*}{UTF8}{gbsn}Y. Sophia Dai (戴昱)\end{CJK*}}
\affiliation{Chinese Academy of Sciences South America Center for Astronomy (CASSACA), National Astronomical Observatories (NAOC), 20A Datun Road, Beijing 100012, China}

\author[0000-0003-2098-9568]{Jonathan P. Gardner}
\affiliation{Astrophysics Science Division, NASA Goddard Space Flight Center, 8800 Greenbelt Rd, Greenbelt, MD 20771, USA}

\author[0000-0003-1530-8713]{Eric Gawiser}
\affiliation{Department of Physics and Astronomy, Rutgers, the State University of New Jersey, Piscataway, NJ 08854, USA}

\author[0000-0003-2775-2002]{Yicheng Guo}
\affiliation{Department of Physics and Astronomy, University of Missouri, Columbia, MO, 65211, USA}

\author[0000-0001-6145-5090]{Nimish Hathi}
\affiliation{Space Telescope Science Institute, Baltimore, MD 21218, USA}

\author[0000-0002-7959-8783]{Pablo Arrabal Haro}
\affiliation{NSF's National Optical-Infrared Astronomy Research Laboratory, 950 N. Cherry Ave., Tucson, AZ 85719, USA}

\author[0000-0001-8587-218X]{Matthew Hayes}
\affiliation{Department of Astronomy and Oskar Klein Centre, Stockholm University, AlbaNova University Centre, SE-10691, Stockholm, Sweden}

\author[0000-0001-9298-3523]{Kartheik G. Iyer}
\affiliation{Dunlap Institute for Astronomy \& Astrophysics, University of Toronto, Toronto, ON M5S 3H4, Canada}

\author[0000-0003-1268-5230]{Rolf A.~Jansen}
\affiliation{School of Earth \& Space Exploration, Arizona State University, Tempe, AZ\,85287-1404, USA}

\author[0000-0001-7673-2257]{Zhiyuan Ji}
\affiliation{University of Massachusetts Amherst, 710 North Pleasant Street, Amherst, MA 01003-9305, USA}

\author[0000-0002-8816-5146]{Peter Kurczynski}
\affiliation{Astrophysics Science Division, NASA Goddard Space Flight Center, 8800 Greenbelt Rd, Greenbelt, MD 20771, USA}

\author[0000-0002-7830-363X]{Maxwell Kuschel}
\affiliation{Minnesota Institute for Astrophysics, University of Minnesota, 116 Church St SE, Minneapolis, MN 55455, USA}

\author[0000-0003-1581-7825]{Ray A. Lucas}
\affiliation{Space Telescope Science Institute, Baltimore, MD 21218, USA}

\author[0000-0002-6016-300X]{Kameswara Mantha}
\affiliation{Minnesota Institute for Astrophysics, University of Minnesota, 116 Church St SE, Minneapolis, MN 55455, USA}

\author[0000-0002-8190-7573]{Robert W. O'Connell}
\affiliation{Department of Astronomy, University of Virginia, Charlottesville, VA 22904-4325, USA}

\author[0000-0002-5269-6527]{Swara Ravindranath}
\affiliation{Space Telescope Science Institute, Baltimore, MD 21218, USA}

\author[0000-0002-4271-0364]{Brant E. Robertson}
\affiliation{Department of Astronomy and Astrophysics, University of California, Santa Cruz, 1156 High Street, Santa Cruz, CA 95064, USA}

\author[0000-0001-7016-5220]{Michael Rutkowski}
\affiliation{Department of Physics and Astronomy, Minnesota State University-Mankato, Mankato, MN, USA}

\author[0000-0002-4935-9511]{Brian Siana}
\affiliation{Department of Physics \&\ Astronomy, University of California, Riverside, CA 92521, USA}

\author[0000-0003-3466-035X]{L. Y. Aaron Yung}
\affiliation{Astrophysics Science Division, NASA Goddard Space Flight Center, 8800 Greenbelt Rd, Greenbelt, MD 20771, USA}

\begin{abstract}
The UltraViolet imaging of the Cosmic Assembly Near-infrared Deep Extragalactic Legacy Survey Fields (UVCANDELS) program provides HST/UVIS F275W imaging for four CANDELS fields. We combine this UV imaging with existing HST/near-IR grism spectroscopy from 3D-HST$+$AGHAST to directly compare the resolved rest-frame UV and \ha\ emission for a sample of 979~galaxies at $0.7<z<1.5$ spanning a range in stellar mass of $10^{8-11.5}$~M$_\odot$. Using a stacking analysis, we perform a resolved comparison between homogenized maps of rest-UV and \ha\ to compute the average \uvha\ luminosity ratio (an indicator of burstiness in star-formation) as a function of galactocentric radius. We find that galaxies below stellar mass of $\sim$10$^{9.5}$ M$_\odot$, at all radii, have a \uvha\ ratio higher than the equilibrium value expected from constant star-formation, indicating a significant contribution from bursty star-formation. Even for galaxies with stellar mass $\gtrsim$10$^{9.5}$~M$_\odot$, the \uvha\ ratio is elevated toward their outskirts ($R/R_{eff}>1.5$), suggesting that bursty star-formation is likely prevalent in the outskirts of even the most massive galaxies but is likely overshadowed by their brighter cores. Furthermore, we present the \uvha\ ratio as a function of galaxy surface brightness, a proxy for stellar mass surface density, and find that regions below $\sim$10$^{7.5}$~M$_\odot$~kpc$^{-2}$ are consistent with bursty star-formation, regardless of their galaxy stellar mass, potentially suggesting that local star-formation is independent of global galaxy properties at the smallest scales. Lastly, we find galaxies at $z>1.1$ to have bursty star-formation regardless of radius or surface brightness.
\end{abstract}


\section{Introduction}
\label{sec:intro}

Star-formation is the basic process by which galaxies assemble their stellar content and is fundamental to the growth of galaxies. Over the past few decades, a significant amount of observational as well as computational effort has been devoted to understanding how star-formation proceeds in galaxies and the physical processes that drive and regulate it. One of the active topics of research continues to be the ``feedback'' processes from stars and central black holes that inject energy and/or momentum into the interstellar medium and prevent or delay the formation of new stars, both in simulations \citep[e.g.,][]{shen14,elbadry16,elbadry17,read19,tacchella22} as well as observationally \citep[e.g.,][]{mcquinn10,sparre17,mercado21}. Understanding the equilibrium between star-formation and its feedback effects is a fundamental challenge for our galaxy formation models, in particular the stellar feedback effects that primarily impact the low-mass scales \citep{springel05,fontanot09,governato10,somerville15,hopkins14}. One of the manifestations of these stellar feedback effects is stochasticity in the star-formation activity, often referred to as ``bursty'' star-formation. Consequently, quantifying this star-formation burstiness from observations of galaxies provides key constraints for theoretical models.

Some spectral features are particularly sensitive to on-going star-formation such as rest-frame UV, IR, submillimeter, and radio continua as well as nebular emission lines (e.g., \ha), which have been used extensively as star-formation indicators. Rest-frame UV and \ha\ are two widely-used star-formation indicators. However, these two tracers are sensitive to star-formation occurring over different timescales: the \ha\ recombination line originates from \hii\ regions powered by the hot O- and B-type stars and decays quickly after $\sim$5~Myrs, whereas the longer lived B- and A-type stars are capable of emitting significant amounts of rest-UV continuum over $\sim$200~Myrs time-scales. Standard calibrations that infer star-formation rates (SFRs) from rest-UV or \ha\ luminosity \citep[e.g.,][]{kennicutt12} assume that those SFRs do not change significantly on $\sim$200~Myr time scales, but approximate an equilibrium state. However, if the SFR varies on timescales shorter than $\sim$200~Myrs, then the rest-UV and \ha\ luminosities also vary in response to the variations in star-formation and in these cases, the standard calibrations break down.

Several studies have used the flux ratio of the rest-UV and \ha\ luminosities \citep[e.g.,][]{glazebrook99,iglesias04,lee09,lee11,weisz12,kaufmann14,dominguez15,mehta17,emami19,fetherolf21} as well as the scatter in the two luminosities \citep[e.g.,][]{broussard19,broussard22,faisst19,griffiths21} to infer the burstiness in star-formation histories (SFHs) of galaxies. There are a number of factors that can affect the \uvha\ ratio such as dust, stellar initial mass functions, stellar metallicity as well as the escape of ionizing photons and the inclusion of binaries and rotating stars in stellar models. However, after accounting for dust, variations in the SFH have the largest impact on the \uvha\ ratio.

Existing observational studies that have used the comparison of rest-UV and \ha\ to infer burstiness in galaxy star-formation have relied on integrated flux measurements and thus are limited to global galaxy properties \citep[e.g., ][]{lee09,weisz12,guo16,emami19}. In this analysis, we use high-resolution imaging as well as slitless spectroscopy from the \textit{Hubble Space Telescope} (HST) to perform a resolved analysis of these two star-formation indicators. The combination of rest-UV imaging from UVCANDELS (X. Wang, et al., in preparation) and near-IR slitless spectroscopy from 3D-HST \citep{brammer12,skelton14,momcheva16} and AGHAST \citep{weiner09} over four of the CANDELS \citep{grogin11,koekemoer11} fields (GOODS-N, GOODS-S, COSMOS and EGS) has provided resolved imaging of both rest-frame UV as well as the \ha\ line emission for galaxies at $0.7<z<1.5$ and thus enabled a unique opportunity for a spatially resolved investigation of burstiness by comparing the two star-formation indicators. In this work, our primary goal is to use the \uvha\ flux ratio to infer the burstiness in galaxy SFHs and quantify its evolution as a function of galaxy structural parameters.

This manuscript is organized as follows: Section~\ref{sec:data} describes the UVCANDELS and 3D-HST$+$AGHAST datasets as well as the selection criteria for the galaxy sample used for this analysis; Section~\ref{sec:maps} describes the data processing performed in order to obtain calibrated and matched maps for the rest-UV and \ha\ emission for our sample; Section~\ref{sec:burstiness_parameter} outlines how the comparison of two star-formation indicators helps us infer details about the SFH; Section~\ref{sec:results} presents our stacking analysis along with the results; Section~\ref{sec:discussion} discusses the impacts of various key assumptions made through the analysis as well as the implications of our results in the context of galaxy SFHs and Section~\ref{sec:conclusions} summarizes our conclusions.

Throughout this paper, we adopt cosmological parameters from Table 3 of \citet{planck}: $\Omega_m=0.315$, $\Omega_\lambda=0.685$ and $H_0=67.31$~km~s$^{-1}$~Mpc$^{-1}$ and all magnitudes used are AB magnitudes \citep{oke83}.

\begin{figure*}[!t]
\centering
\includegraphics[width=0.72\textwidth]{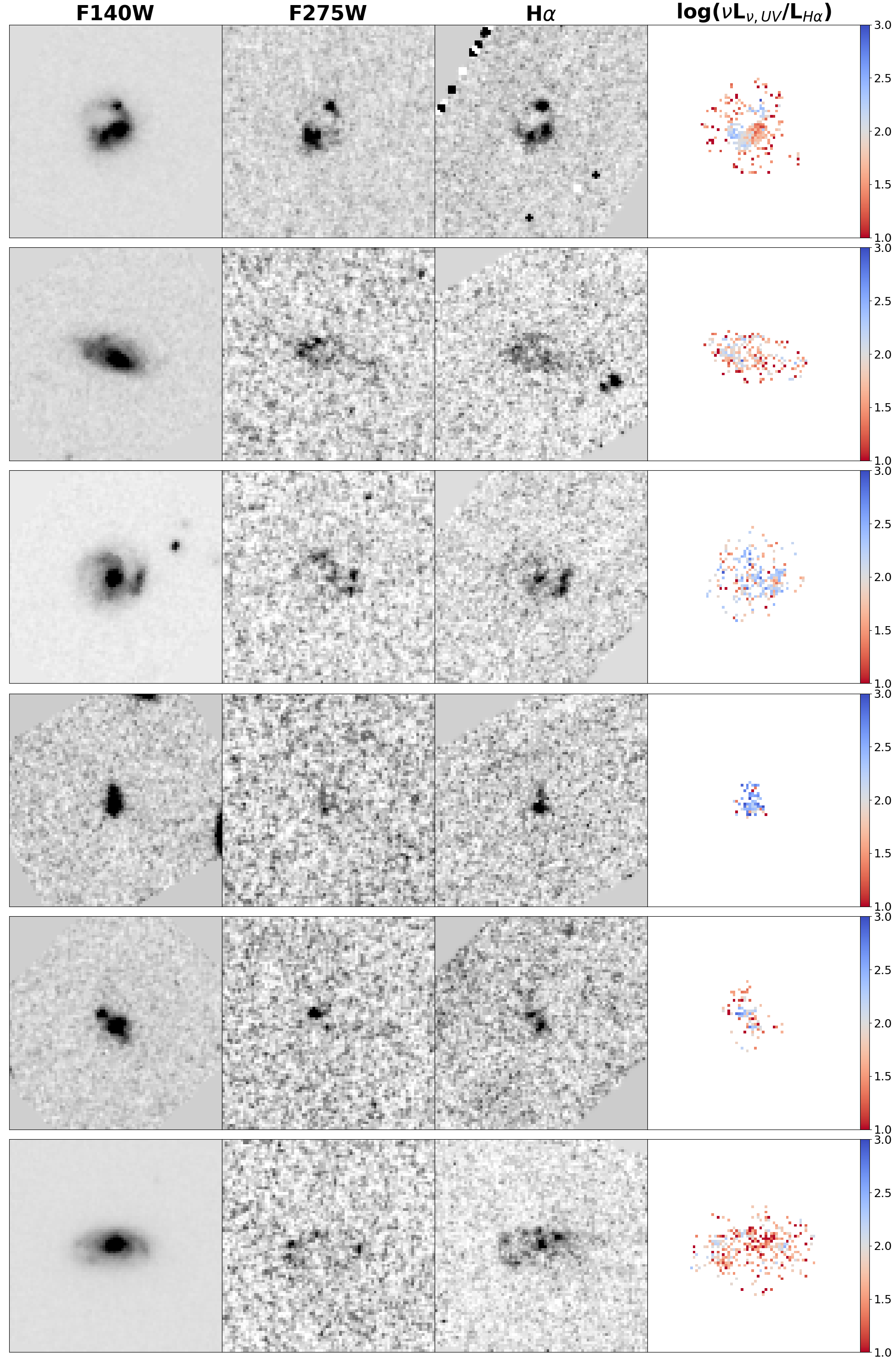}
\caption{A few hand-picked examples of the extended galaxies in our sample with their (\textit{from left to right}) F140W, rest-frame UV, \ha, and \uvha\ ratio maps. While the majority of our sample consists of objects that are more compact, these selected examples explicitly highlight the spatial variation between the rest-UV and \ha\ emission. The stamps shown are all 8\arcsec$\times$8\arcsec\ (at 0.1\arcsec\ pixel-scale). See Section~\ref{sec:maps} for details on how the maps are generated. Some of the bright features (not associated with the galaxy) visible in the \ha\ stamps are either emission lines from chance alignment with spectral traces from other nearby sources in the grism observations, hot pixels, or artifacts from data reduction in the \ha\ stamps that typically appear at the edge of the grism cutout. The \uvha\ ratio maps for individual galaxies are shown here purely as a visual aid; our analysis and results rely on first stacking the rest-UV and \ha\ emission (see Section~\ref{sec:stacking} for details).}
\label{fig:poster_examples}
\end{figure*}

\section{Data and Sample selection}
\label{sec:data}
The galaxy sample used for the analysis presented in this work is obtained from the recently completed UVCANDELS Hubble Treasury program.

\subsection{UVCANDELS Imaging}
\label{sec:data_uvcandels}
The Ultraviolet Imaging of the Cosmic Assembly Near-infrared Deep Extragalactic Legacy Survey Fields (UVCANDELS; PI: H. Teplitz; PID: 15647; X. Wang, et al., in preparation) is a Cycle~26 Hubble Treasury program that obtained UV imaging of four of the deep-wide survey fields defined by CANDELS \citep{grogin11,koekemoer11}. UVCANDELS consists of WFC3/F275W imaging over a total of $\sim$430~arcmin$^2$ combined area over four CANDELS fields (GOODS-N, GOODS-S, COSMOS, and EGS) reaching an imaging depth of AB=27 for compact galaxies, along with ACS/F435W (reaching AB$\ge$28.0) imaging in parallel. The UVCANDELS F275W mosaics are available at the Mikulski Archive for Space Telescopes (MAST; \dataset[doi:10.17909/8s31-f778]{https://dx.doi.org/10.17909/8s31-f778})\footnote{\url{https://archive.stsci.edu/hlsp/uvcandels}} for all four fields along with F435W mosaics for COSMOS and EGS\footnote{The F435W mosaics for GOODS-N and GOODS-S are already available from CANDELS.}. These UVCANDELS image mosaics are aligned and registered to the CANDELS astrometry. With the newly acquired UVCANDELS imaging, we have measured the fluxes in the F275W filter for the four fields, along with the deeper F435W for COSMOS and EGS. We adopt a similar methodology as \citet{rafelski15} where F275W flux measurements rely on isophotes defined on the F606W mosaics optimizing the signal-to-noise for the NUV photometry of small faint sources followed by an aperture correction to match it to the $H$-band isophote-based photometry for the optical and near-IR bands. The full description of the UVCANDELS dataset including the mosaics as well as photometry will be provided in X. Wang, et al. (in preparation). The stellar physical properties for these galaxies as estimated from spectral energy distribution (SED) fitting will be available from V. Mehta, et al. (in preparation; but see Section~\ref{sec:sample} for a brief description).

The galaxy sample used in the analysis presented in this work is selected from the UVCANDELS photometric catalogs. The full sample selection is described in Section~\ref{sec:sample}. We also use the image mosaics generated from UVCANDELS for our resolved analysis since the F275W filter allows us to directly probe the rest-frame UV light from galaxies over our redshift range of interest, $0.7<z<1.5$. Furthermore, we also use the F435W mosaics\footnote{We use the mosaics from UVCANDELS for COSMOS and EGS fields, and the existing ones from CANDELS for GOODS-N and GOODS-S fields.} for deriving resolved dust maps (see Section~\ref{sec:maps_dust}) as well as the F160W mosaics from CANDELS\footnote{\url{https://archive.stsci.edu/hlsp/candels}} to ensure accurate astrometric alignment (see Section~\ref{sec:maps_alignment}).

\subsection{3D-HST$+$AGHAST Grism Spectroscopy}
\label{sec:data_3dhst}
We use the archival slitless grism spectroscopy data from the 3D-HST \citep{brammer12,skelton14,momcheva16} and A Grism H-Alpha SpecTroscopic survey \citep[AGHAST,][]{weiner09} surveys for measuring the \ha\ emission for our galaxy sample. The 3D-HST$+$AGHAST program is a wide-field, near-infrared grism spectroscopy HST-based survey that obtained WFC3/G141 grism coverage for $\sim$500~arcmin$^2$ over all five CANDELS fields. In this work, we use the recent re-analysis of this archival HST grism data in the CANDELS fields carried out as part of the Complete Hubble Archive for Galaxy Evolution (CHArGE) initiative \citep[see][]{kokorev22}.

Briefly, this reduction of 3D-HST$+$AGHAST data uses the state-of-the-art \grizli\footnote{\url{https://github.com/gbrammer/grizli}} software \citep{brammer21}, which executes an iterative forward-modeling approach to perform quantitative fitting of the spectral trace for each object in the field-of-view of the grism exposures at the visit level. This approach also includes a comprehensive modeling of contamination due to nearby objects and their own spectral traces. As its final science-ready data products, \grizli\ provides cutouts of the direct image and segmentation maps used for modeling the object as well as the 1D/2D grism spectra and continuum-subtracted emission line maps \citep[see e.g.,][]{wang22}. In this analysis, we use the \ha\ emission line maps as well as a cutout of the F140W mosaic which serves as a direct image for the grism spectral extraction.

\subsection{Sample Selection}
\label{sec:sample}
The galaxy sample for our resolved analysis of rest-frame UV and \ha\ emission is selected to include all galaxies with a significant detection in the UVCANDELS F275W images as well as in their measured \ha\ fluxes. Specifically, we require the signal-to-noise ratio ($SNR$) to be $>5$ for both the measured integrated F275W flux and the integrated \ha\ emission-line flux. The wavelength coverage of the G141 grism (1.1$-$1.7\micron) establishes the redshift range for our sample as $0.7<z<1.5$, where it covers the rest-frame \ha. Over this redshift range, the F275W filter probes the rest-frame $\sim1100-1600$\AA\ light, thus giving access to the second SF indicator, the rest-UV.

Before finalizing our galaxy sample, we remove any objects that do not have F435W coverage\footnote{The coverage in F435W has some gaps due to the nature of the parallel observation setup for simultaneous F275W and F435W imaging. This is a geometrical selection that does not introduce biases for our analysis.} which is required for our dust measurement (see Section~\ref{sec:maps_dust} for details). We also exclude any potential AGN candidates from the deep Chandra X-ray observations in these CANDELS fields \citep[][D. Kocevski, priv. comm.]{hsu14, nandra15, xue16}. Furthermore, we perform a round of visual vetting to exclude any objects that are affected by data quality issues or incorrect redshift estimate due to line mis-identification, source confusion, trace contamination or other artifacts. While the \ha\ and \nii\ doublet are not resolved in the G141 grism spectra, the other emission feature that is most common for our sample, i.e., \hb+\oiii\ complex, is marginally resolved and is easily identified. Our visual vetting step serves as an additional check to ensure the sample purity; however, the \grizli\ redshift estimate from the combined fit to the grism spectrum and photometry is robust and we do not find any objects with misidentified emission lines.

Our final sample consists of 979~galaxies in the redshift range of $0.7<z<1.5$ over four CANDELS fields (GOODS-N, GOODS-S, COSMOS, and EGS). Figure~\ref{fig:poster_examples} shows a few examples hand-picked to have high $SNR$ as well as extended morphologies in order to demonstrate the variation between the rest-UV and \ha\ emission. These examples are not representative of our full galaxy sample as a majority of the objects are rather compact and not as morphologically impressive.

\subsection{Estimating the stellar physical properties}

\begin{figure}[!t]
\centering
\includegraphics[width=0.48\textwidth]{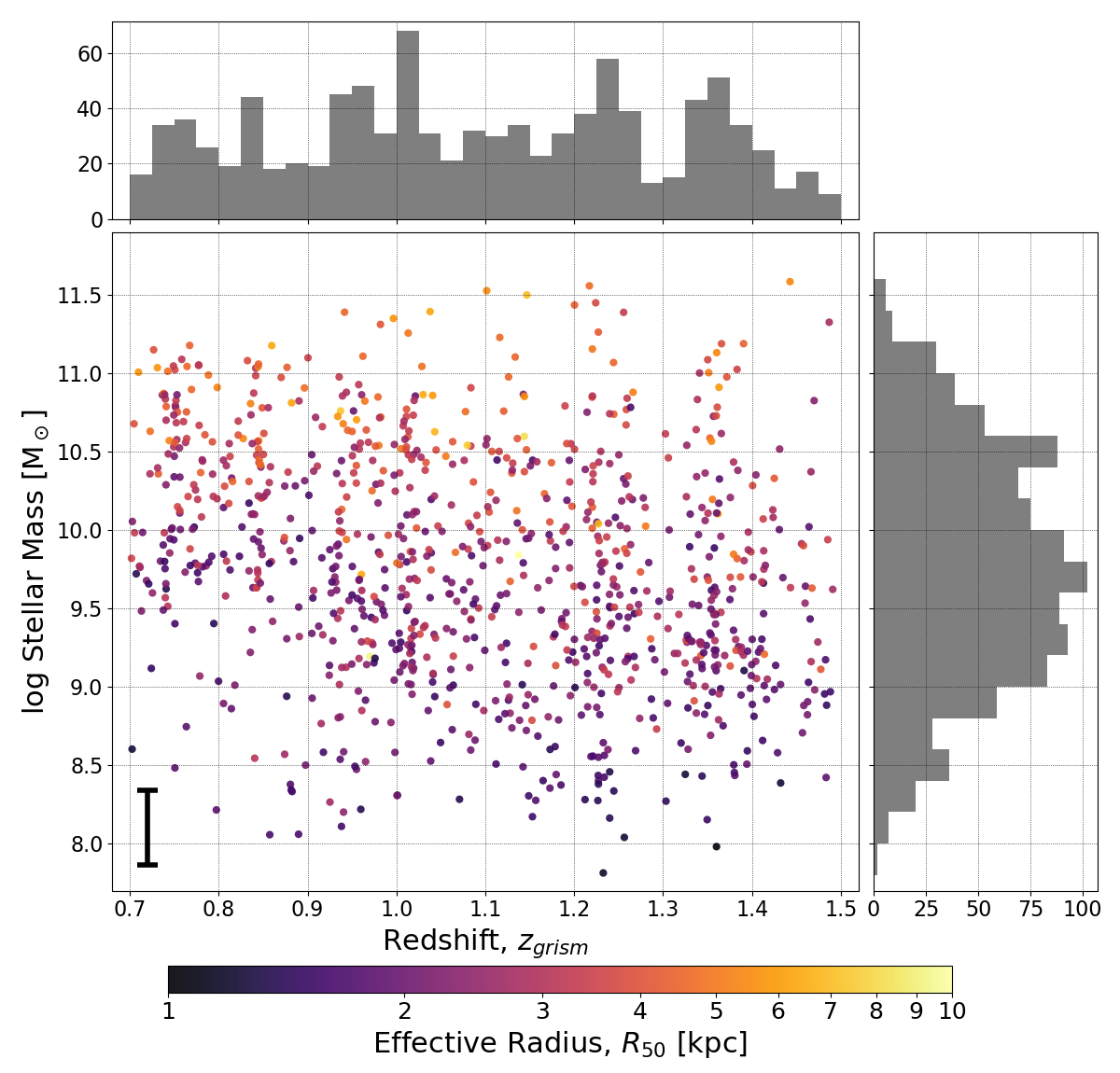}
\caption{The redshift and stellar mass distribution of our final sample of 979~galaxies over $0.7<z<1.5$ with rest-UV coverage available from UVCANDELS and \ha\ from 3D-HST$+$AGHAST. The symbols are colored according to their effective radius ($R_{eff,50}$) as measured in the F160W filter (as reported in the CANDELS photometric catalogs), converted to physical units. The typical 1$\sigma$ error-bar for the best-fit stellar masses is shown in the \textit{lower left corner}.}
\label{fig:sample_summary}
\end{figure}

The spectral information from 3D-HST$+$AGHAST grism data provides precise redshifts for our sample and the stellar physical properties are computed using these grism redshifts. Briefly, the stellar population properties are measured on integrated photometry using the \dbasis\ code \citep{dbasis}, which is based on the Flexible Stellar Population Synthesis library \citep[FSPS;][]{conroy09,conroy10} and utilizes Gaussian processes to parameterize flexible SFHs. We define the \dbasis\ template grid with a set of free parameters: stellar metallicity ($\log{(Z/Z_\odot)}$ varying over $-1.5$ to $0.25$), dust ($A_V$ varying from 0 to 4), and four ``shape'' ($t_X$) parameters to define the SFHs. We assume \citet{chabrier03} IMF and \citet{calzetti00} dust law for the model templates. \dbasis\ outputs the model template that best fits the observed data along with the estimated stellar mass, SFR and other stellar population properties. Figure~\ref{fig:sample_summary} shows the redshift and stellar mass distributions for our final sample. These distributions illustrate that our sample uniformly spans the redshift range $0.7<z<1.5$. The sample includes galaxies that range in mass from $10^{8-11.5}$~M$_\odot$, roughly corresponding to the present day Small Magellanic Cloud (SMC) at the low mass end to the Milky Way at the high mass end.

\section{Resolved maps of rest-frame UV and \ha\ emission}
\label{sec:maps}

We start with the final data products available from the \grizli\ 3D-HST$+$AGHAST reduction, which includes a 8\arcsec$\times$8\arcsec\ (at 0.1\arcsec\ pixel scale) continuum-subtracted emission map for all detected lines and image cutout in the direct images (typically, F140W). We then extract a matching cutout from the UVCANDELS F275W mosaic (at 0.06\arcsec\ pixel scale) using \swarp\footnote{\url{https://github.com/astromatic/swarp}}, which allows us to adjust the pixel scale as well as image orientation to match that of the \ha\ maps from \grizli.

\subsection{Aligning UVCANDELS and 3D-HST$+$AGHAST}
\label{sec:maps_alignment}
The astrometric reference frames used to align the UVCANDELS mosaics and the \grizli\ reduction of the 3D-HST$+$AGHAST data, which are used for this work, are not consistent. Since we are primarily interested in object-specific cutouts for our analysis, we need to account for any mis-alignment on an object-by-object basis when making their cutouts.

We measure the astrometric offset from a cross-correlation of the images. In order to minimize the impact of galaxy morphology at different wavelengths, we use the CANDELS F160W mosaic which has an identical astrometric reference as UVCANDELS and is the closest match in wavelength coverage to the F140W direct images from 3D-HST$+$AGHAST. Additionally, we mask all objects other than the target galaxy to ensure there are no mismatches. Finally, we spatially cross-correlate the CANDELS F160W cutout with the 3D-HST$+$AGHAST F140W from \grizli\ to quantify the astrometric offset between the two datasets. The astrometric offsets vary from field to field but are within $\sim$0.2\arcsec\ in R.A. and within $\sim$0.1\arcsec\ in Decl. for our sample. We apply the corrections necessary for proper alignment when extracting the final cutouts used for further analysis.

\subsection{Calibrating the maps}
\label{sec:maps_calibration}
The aligned and matched F275W and \ha\ emission maps are corrected for residual background by performing a local background estimation and subtracting it from the stamps. The maps are then calibrated to account for instrument and filter response and converted into luminosity units (\ergsAkpc\ for rest-UV; and \ergskpc\ for \ha).

The maps are further corrected for galactic extinction using the \citet{schlegel98} dust maps\footnote{Specifically, we use the Python implementation available at \url{https://github.com/adrn/SFD} to query the SFD dust maps.}. The reddening $E(B-V)$ is queried at the location of each galaxy and converted to an extinction assuming a \citet{cardelli89} extinction law, for the respective filters as well as for \ha\ at its observed wavelength.

Since the \ha\ and \nii$\lambda\lambda6548+6584$ doublet are blended at the G141 resolution, we need to apply a correction to the \ha\ emission maps to remove the \nii\ contribution. The \nii/\ha\ ratio in galaxies has been shown to vary by up to factors of $3-5$ and this variation has been linked to various galaxy properties such as its metallicity, stellar mass, SFR, and nitrogen-to-oxygen ratio ($N/O$) as well as known to evolve as a function of redshift \citep[e.g.,][]{kewley13,masters14,steidel14,shapley15,masters16,strom17,faisst18}. Here, we use the mass- and redshift-dependent calibration from \citet{faisst18} to apply the appropriate correction for each galaxy according to its redshift and best-fit stellar mass.

\subsection{Correcting for dust}
\label{sec:maps_dust}

Before being interpreted as SF indicators, both rest-UV and \ha\ emission need to be corrected for attenuation due to the presence of dust in the host galaxy itself. Since the main goal of this work is to compare the rest-UV and \ha\ emission from galaxies as a function of their position inside the galaxy, it is imperative to make an effort to address the fact that dust absorption may itself vary inside a galaxy \citep[e.g.,][]{kim19}. Hence, we compute a resolved dust map for each galaxy to correct for dust attenuation while preserving the morphological detail.

In order to compute the resolved dust map, we use the relation between the UV slope \citep[$\beta$;][]{calzetti94} and the dust attenuation ($A_V$). We rely on imaging in the F275W, F435W and F606W filters in order to constrain $\beta$. This process is less precise than SED fitting, but performing a (pixel-by-pixel) resolved SED fit is beyond the scope of this work. We estimate $\beta$ assuming the functional form $f_\lambda \propto \lambda^\beta$ using photometry from the filters closest to the rest-frame near-UV. For objects with redshifts $z\lesssim1.08$, we use the F275W and F435W filters; whereas, for objects with $z\gtrsim1.08$, where the F275W filter falls below rest-frame 1300\angstrom, we use the F435W and F606W filters\footnote{where F606W imaging is available}.

The measurement of $\beta$ with this methodology is unreliable for low-$SNR$ pixels. For each galaxy where the individual pixels fall below a $SNR$ of 3, we instead use a default value for $\beta$ that is computed from the total integrated flux for the galaxy in the relevant filters. This ensures that $\beta$ is measured consistently and reliably for all pixels. As a sanity check, we confirm that the UV slopes measured from the two-band integrated photometry for the full galaxy are in good agreement with those computed from the best-fit SED models.

\begin{figure}[!t]
\centering
\includegraphics[width=0.4\textwidth]{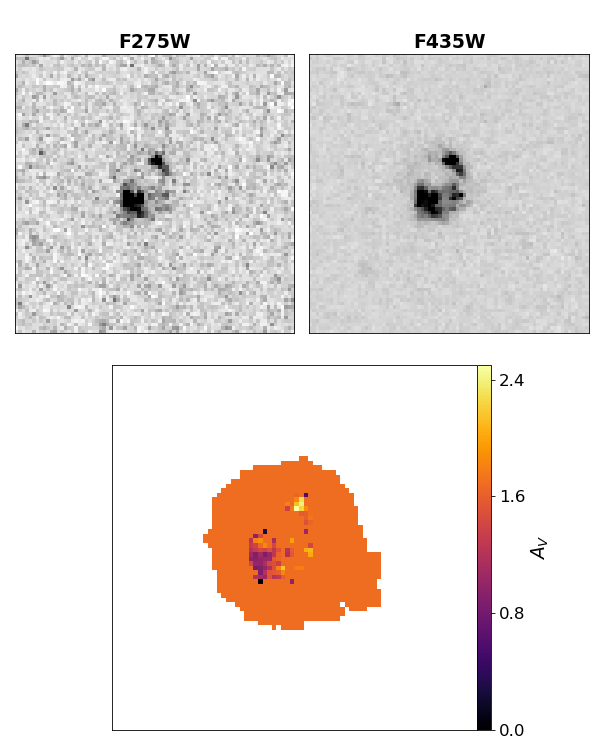}
\caption{The resolved dust map for an example galaxy computed via the UV slope $\beta$ estimated from the F275W and F435W images and the \citet{reddy18} calibration of the the $\beta-A_V$ relation (see text for details). The final dust map consists of independent $A_V$ measurements for individual pixels with sufficient $SNR$ and a default value computed from the total integrated galaxy flux in F275W and F435W for pixels with $SNR<3$.}
\label{fig:dust_example}
\end{figure}

With the maps of UV slope $\beta$ computed for our sample, we then use the $\beta$--$A_V$ relation to convert to a resolved map of dust attenuation, $A_V$. Specifically, we use the calibration from \citet{reddy18} assuming the \citet{calzetti00} dust law\footnote{$\beta=-2.616 + 4.684 \times E(B-V)$}. We confirm that only 12\%, 9\%, 4\%\ of the per-pixel $\beta$ values and only 0.2\%\ of the $\beta$ values computed from integrated galaxy photometry fall below the dust-free $\beta_0=-2.616$ value implied for the \citet{reddy18} model galaxy templates. Given that measurement noise, particularly for low SN pixels, is likely to drive the extended tails of the $\beta$ distribution, we believe the models from \citet{reddy18} (with $\beta_0=-2.616$) is not inappropriate for our sample. Figure~\ref{fig:dust_example} shows an example of the resolved dust map generated following the steps outlined here. Lastly, we use this $A_V$ map to apply a dust correction to the rest-UV and \ha\ maps based on the \citet{calzetti00} dust law, assuming $E(B-V)_{stellar}/E(B-V)_{nebular} = 0.44$ \citep{calzetti00}. For the values computed from the integrated galaxy flux, the median $\beta$ for our sample is $-1.17$ with an inter-quantile range (IQR) spanning over $-1.53$ to $-0.78$, where as the median $E(B-V)_{stellar}$ value is $0.3$ with an IQR of $0.22$ to $0.38$.

\subsection{Matching the PSF}
The point-spread function (PSF) evolves considerably between the near-infrared (G141 which covers \ha) and F275W (which provides rest-UV). In order to be able to draw accurate comparisons between the spatial extent of rest-UV and \ha, it is imperative to correct for the change in the PSF. We address this issue by convolving the F275W maps to match the near-infrared PSF. The target PSF is selected from WFC3/IR F105W, F125W, and F160W, whichever is the closest match (in wavelength) to the observed wavelength of the \ha\ emission. We then use \pypher\footnote{\url{https://pypher.readthedocs.io/}} \citep{pypher} to compute a convolution kernel that, when applied, converts the F275W PSF to that of the \ha\ emission.

\section{Using rest-UV and \ha\ to infer burstiness}
\label{sec:burstiness_parameter}

\begin{figure*}[!t]
\centering
\includegraphics[width=0.48\textwidth]{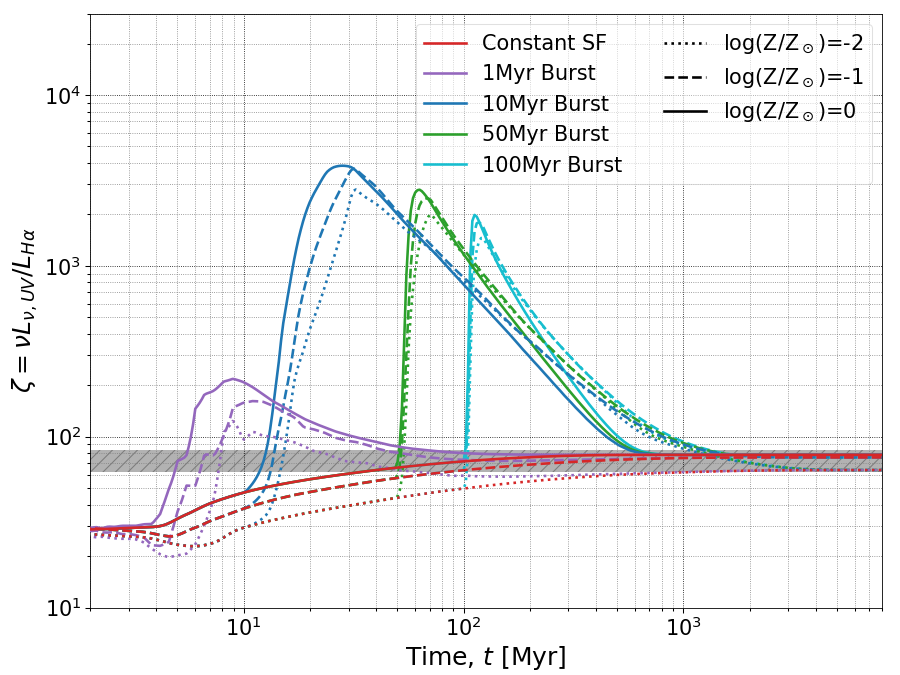}
\includegraphics[width=0.48\textwidth]{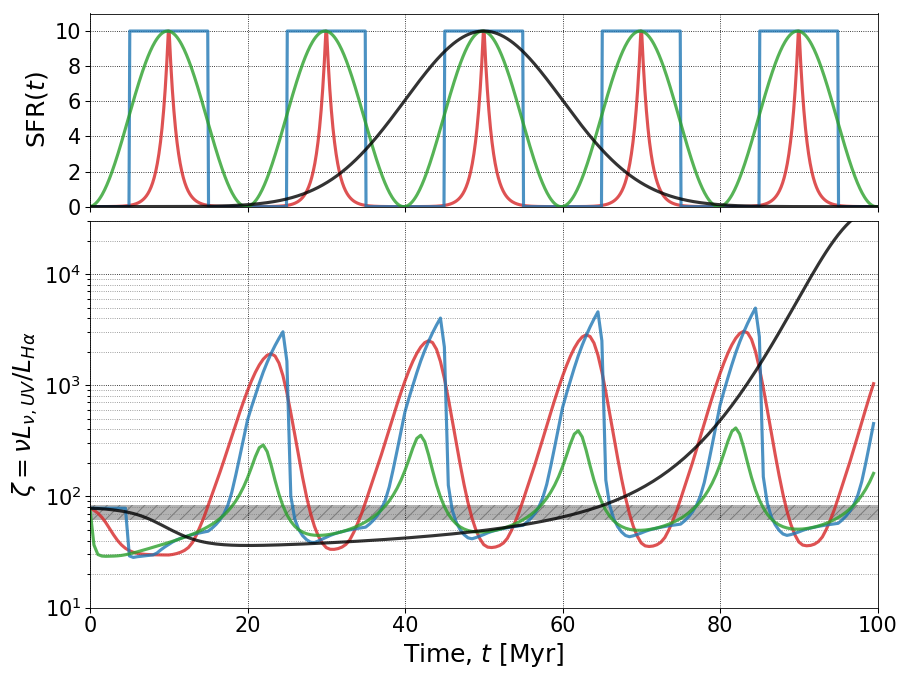}
\caption{The time-evolution of the \textit{dust-corrected} \uvha\ ratio ($\zeta$) as computed from stellar population synthesis models (see text for details on the models). \textit{Left panel}: An idealistic scenario showing SFHs with a single burst of varying duration (1, 10, 50, 100~Myrs shown in \textit{purple, blue, green, cyan} curves, respectively) atop a low-level of underlying star-formation. In constant star-formation case (\textit{red} curve), the value of $\zeta$ reaches an equilibrium value (shown as the \textit{grey hatched, shaded} band) within $\sim$200~Myr, where as in the burst cases, the value of $\zeta$ quickly deviates significantly from its equilibrium state after the burst ends. \textit{Right panel}: An idealistic scenario with showing the behavior of $\zeta$ in the case of multiple bursts of varying shapes. The value of $\zeta$ is driven below that of the equilibrium state (\textit{grey hatched, shaded} region; same as in the \textit{left panel}) during the rising phase of the burst, and above the equilibrium during the declining phase. The SFHs with sharper evolution (top-hat in \textit{blue} and exponentially-peaked in \text{red}) result in a sharper change in $\zeta$ compared to the smoother sinusoidal (in \textit{green}) bursts. On the other hand, a long-term smooth rise/decline in SFR (shown in \textit{black}) is also able to reproduce similar deviations in the $\zeta$ value as the short-term bursts.}
\label{fig:uvha_templates}
\end{figure*}

The rest-frame UV continuum and the \ha\ recombination line are both indicators of on-going star-formation activity in galaxies \citep{kennicutt94, kennicutt12}. However, these indicators are sensitive to star-formation activity over different timescales. The \ha\ recombination line originates in the \hii\ regions surrounding the hot O- and B-type stars and it decays rapidly $\sim$5~Myr after a cessation of star formation. On the other hand, the longer-lived B- and A-type stars continue to power the total rest-frame UV continuum over a $\sim$200~Myr timescale. This difference in the timescales traced by rest-UV and \ha\ enables us to probe the star-formation activity in galaxies over the recent $\sim$200~Myr.

Several studies have performed comparisons across these two star-formation indicators to infer the intrinsic properties of the galaxy SFHs in observational surveys \citep[e.g.,][]{glazebrook99,iglesias04,boselli09,lee09,lee11,lee12,weisz12,guo16,mehta17,emami19,broussard22} as well as in simulations \citep{shen13,hopkins14,dominguez15,broussard19,emami21,griffiths21}. These studies have typically compare the ratio of the luminosities of the two indicators or its scatter to infer the galaxy SFH. However, there are several other effects that also affect the ratio of the observed rest-UV and \ha\ luminosities from galaxies:
\begin{itemize}
    \item Dust Attenuation: The attenuation of a galaxy spectrum due to dust is strongly dependent on wavelength. Moreover, the spatial distribution of the dust can be non-uniform, where the dust content attenuating the light from the stellar birth clouds is different compared to that attenuating the nebular emission \citep{kewley02,lee09}.

    \item Initial Mass Function (IMF): The ratio of rest-UV and \ha\ luminosities is sensitive to the relative number of stars at each mass and hence is affected by the IMF. There are two main scenarios to consider, both of which manifest at small (low-SFR) scales: (\textit{i}) A non-universal IMF such as the integrated galactic IMF \citep[IGIMF,][]{weidner05,pflamm07,pflamm09,weidner11} could lead to a steeper IMF slope at low-SFR values \citep[$<10^{-2}$~M$_\odot$~yr$^{-1}$; ][]{pflamm09} due to a lower chance of forming mass stars relative to the high-SFR case. (\textit{ii}) A universal IMF may not fully sampled at all SFRs and could be biased against the very massive stars. Low-SFR galaxies may not be able to properly sample the high-mass end of the IMF leading to stochasticity on a star cluster scale \citep[e.g.,][]{fumagalli11,eldridge12,dasilva12,cervino13}. However, the impact from this stochastic IMF sampling, by itself, has been shown to be smaller compared to that of the IGIMF case \citep{fumagalli11,andrews13,andrews14}.

    \item Stellar Metallicity: The stellar abundance affects the stars' temperatures, where stars with lower metal abundances are hotter, and thus, the \ha\ emission is intensified relative to the UV resulting in a lower value of the \uvha\ ratio compared to their higher metallicity counterparts; however, this effect is minimial \citep[$\sim$10\%;][]{bicker05,boselli09}.

    \item Other effects such as the escape of ionizing photons as well as the inclusion of binary \citep{eldridge12} and rotating stars \citep{choi17} in modeling stellar populations can also impact the \uvha\ ratio.
\end{itemize}

While all of these factors affect the observed ratio of the rest-UV and \ha\ luminosities, the impact of dust attenuation and the SFH is the most dominant. However, attenuation due to dust is well-parameterized with the help of dust laws (e.g., \citealt{calzetti00}) and the dust content of galaxies can be measured independently. Hence, we can correct the ratio of the observed rest-UV and \ha\ luminosities for dust and use it to infer the SFHs of galaxies. For the sake of convenience, we introduce a \textit{``burstiness''} parameter ($\zeta$) in the rest of our analysis, which is defined as the \textit{dust-corrected} \uvha\ luminosity ratio:
\begin{equation}
    \zeta = \log{(\nu L_{\nu,UV} / L_{H\alpha})}
\end{equation}

Figure~\ref{fig:uvha_templates} demonstrates the evolution of $\zeta$ under different assumptions of SFHs, stellar IMFs and metallicities using theoretical galaxy model templates. We use the Flexible Stellar Population Synthesis library \citep[FSPS;][]{conroy09,conroy10} with the MILES stellar library and MIST\footnote{\url{http://waps.cfa.harvard.edu/MIST/}} isochrones to generate the spectral energy distribution of our mock stellar populations. We compute the rest-UV (1500\angstrom) luminosity density and \ha\ line luminosities for a range of stellar populations to track the evolution of $\zeta$ over a 10~Gyr timescale. We vary the IMF for the stellar population amongst \citet{chabrier03}, \citet{salpeter55}, and \citet{kroupa01}. The stellar metallicity is varied over a range of $log{(Z/Z_\odot)}$=$-2$ to 0 and the gas-phase metallicity is set to be equal to the stellar. We compare star-formation bursts of varying durations and burst shapes with the constant star-formation scenario. In our toy model, the star-bursts are introduced on top of a low level of underlying constant star-formation at the level of $\sim$0.1\%\ of the peak burst amplitude, which is consistent with the SFHs inferred for $\sim$10$^{8-10}$~M$_\odot$ galaxies observed in the local universe \citep[e.g., ][]{weisz12,kaufmann14,emami19}. Lastly, since we are considering the \textit{dust-free} luminosity ratio ($\zeta$), there is no dust added to our toy model.

The evolution of $\zeta$ in an idealistic scenario of individual bursts of star-formation is shown in the left panel of Figure~\ref{fig:uvha_templates}. In the case of constant star-formation, the value of $\zeta$ asymptotes to an equilibrium value after $\sim100-200$~Myr. Accounting for variations in IMF and stellar metallicities, the equilibrium value for $\zeta$ spans a narrow range of $\sim$60 to 85. However, in the case of bursts, $\zeta$ quickly deviates from its constant star-formation value both during the rise and fall of star-formation. Initially, during the rise, $\zeta$ decreases due to the increasingly bright \ha\ emission from the hot O- and B-type stars. Toward the end of the burst, once the star-formation starts to decline the value of $\zeta$ increases significantly (by several dex) as the O- and B-type stars die off and the \ha\ emission starts to decay, while the rest-UV continuum survives for a longer duration. Ultimately, the strength of both the rest-UV and \ha\ emission from the star-burst decay to the point where their contribution relative to low-level, underlying constant star-formation is not significant and the value of $\zeta$ returns to its equilibrium value. The general trend in the time-evolution of $\zeta$ is reproduced regardless of the metallicities. Variations due to the choice of fiducial IMFs are on the scale of the line-width of the curves drawn in Figure~\ref{fig:uvha_templates} (left panel), and are therefore negligible. Varying the upper mass cutoff (M$_{up}$) for the IMF can mimic a increase in the value of $\zeta$. In the case of \citet{chabrier03} IMF, changing M$_{up}$ from its default value (100~M$_\odot$) to 80, 60, 40, 20 M$_\odot$, causes the equilibrium $\zeta$ value to rise from 90 to 100, 140, 240, 1640, respectively. However, it is important to caution that tweaking the M$_{up}$ value in this way is rather arbitrary; instead, the IGIMF provides a more physically-motivated formulation of this effect.

The right panel of Figure~\ref{fig:uvha_templates} instead illustrates the behavior of $\zeta$ in the case of periodic short bursts that are again introduced atop a low, underlying baseline. The stellar IMF is fixed to be \citet{chabrier03} and the stellar metallicity is fixed to solar for this case. The value of $\zeta$ responds to the individual bursts similar to the earlier case, but its time-evolution is now truncated by the onset of new star-bursts, which dominate the relative contribution to the intrinsic rest-UV and \ha\ luminosities. The deviation in $\zeta$ from the equilibrium value retains its asymmetrical behavior, where $\zeta$ deviates significantly more (by up to several dex depending on the strength of the burst) after a burst compared to during the burst onset/rise. Thus, provided that the star-formation is not constant and the duration of bursts is on the order of the timescales between two consecutive bursts or shorter, the observed value of $\zeta$, on average, will be systematically higher than the equilibrium value for constant star-formation. The exact shape of the burst has minimal impact on the behavior of $\zeta$ as evident from Figure~\ref{fig:uvha_templates}. The change in $\zeta$ is only slightly exaggerated in the case of sharply changing SFHs (the exponentially-peaking and top-hat cases), compared to the sinusoidal case.

However, it is important to mention that smoothly rising or falling SFHs can also result a similar behavior in $\zeta$. A long-term smooth decline (rise) in SFR also causes $\zeta$ to be systematically higher (lower) than the equilibrium value as illustrated in Figure~\ref{fig:uvha_templates} (right panel). This case is similar to having a single smoothly evolving burst and the rate of change in SFR is the primary factor that controls the deviation in $\zeta$. Nonetheless, long-term smoothly rising/falling SFHs can also be responsible for the observed $\zeta$ value, similar to the case with multiple short-term bursts.

\begin{figure*}[!ht]
\centering
\includegraphics[width=0.38\textwidth]{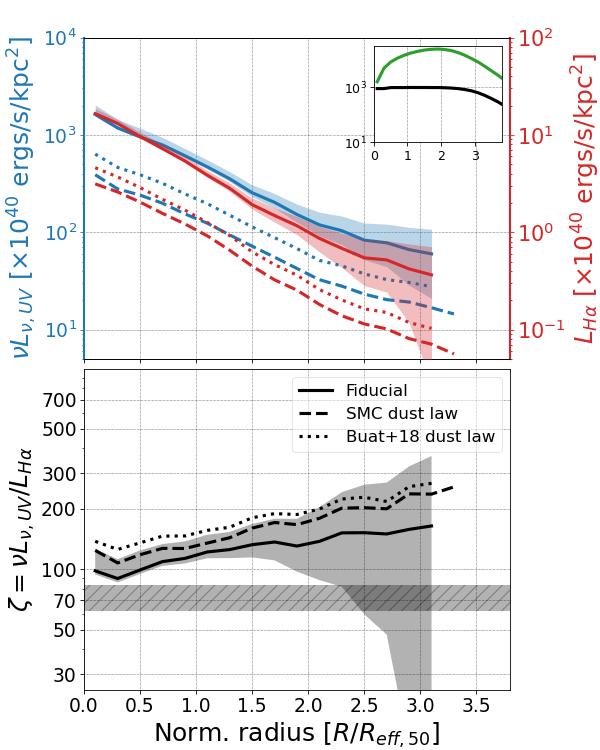}
\hspace{0.25in}
\includegraphics[width=0.38\textwidth]{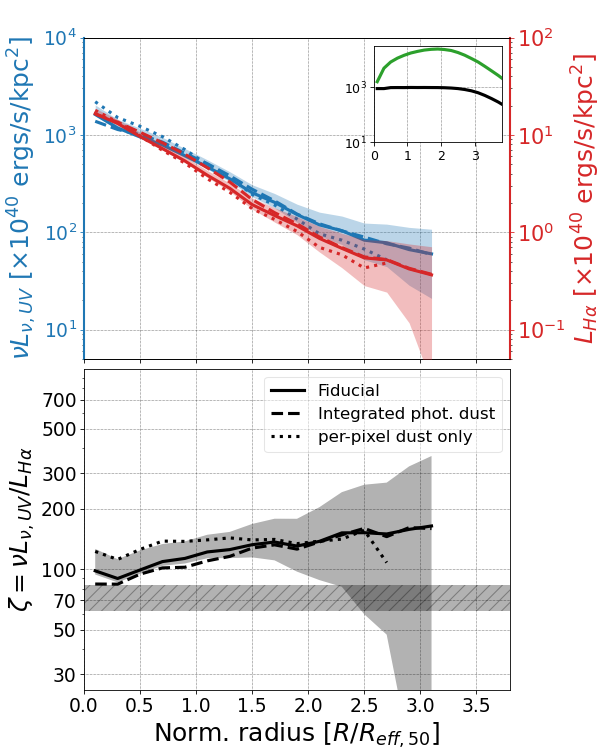}
\caption{The average light profile as function of the galactocentric radius for rest-UV and \ha\ emission (shown in the \textit{top panels}) as well as the \uvha\ ratio ($\zeta$; shown in the \textit{bottom panels}) computed for our galaxy sample (see text for details). The \textit{grey hatched, shaded} band shows the equilibrium value expected for constant star-formation case. In addition to our fiducial case (shown in \textit{solid} curves) which uses resolved dust maps and the \citet{calzetti00} dust law, we also show additional cases for comparison: In the \textit{left panel}, \textit{dashed} curves show the result assuming the SMC \citep{gordon03} dust law and the \textit{dotted} curves show that assuming the \citet{buat18} dust law instead; and in the \textit{right panel}, \textit{dashed} curves show the result if we assumed a uniform dust value for each galaxy; and \textit{dotted} curves show the result if we only consider the pixels with a reliable per-pixel dust estimate. The \textit{inset} shows the total number of galaxies (in \textit{black}) and pixels (in \textit{green}) that fall within each individual bin in radius. The \textit{shaded} bands show the 1$\sigma$ error on the light profiles.}
\label{fig:result_reff_extended}
\end{figure*}

\section{Results}
\label{sec:results}

\subsection{Rest-UV and \ha\ light profiles via stacking}
\label{sec:stacking}

The main goal of this work is to study the spatially-resolved evolution of burstiness in galaxy SFHs. With the resolved maps of rest-UV and \ha\ emission in hand (from Section~\ref{sec:maps}), we now focus on computing the burstiness parameter ($\zeta$, Section~\ref{sec:burstiness_parameter}) as a function of galaxy structural parameters, namely galactocentric radius and stellar mass surface density. On an individual galaxy basis, variations in $\zeta$ can be rather large due to several galaxy property-dependent factors. We marginalize over these to calculate an average value of $\zeta$, which can then provide an insight on the average properties of the general galaxy population. In this work, we adopt a stacking approach to compute an azimuthally-averaged $\zeta$ for our sample of 979~galaxies.

To do this, we start with the rest-UV and \ha\ maps computed from Section~\ref{sec:maps} that have been properly aligned, calibrated, corrected for dust and matched for PSF variations. Furthermore, we also process the F140W image stamps from 3D-HST$+$AGHAST for each galaxy through the same steps. We use the F140W stamps, which trace the rest-frame optical light for our sample, to define a centroid that is not as strongly affected by recent star-formation activity and gives a more robust galaxy centroid. For each galaxy, we then bin all pixels belonging to the galaxy\footnote{For rest-UV, the segmentation map is defined using the F435W images as per the UVCANDELS photometry (X. Wang, et al., in preparation); whereas for \ha, it is defined as per \grizli\ using the F140W direct imaging that was obtained alongside the grism images.} according to their galactocentric radius. We define the galactocentric radius as the distance from the centroid normalized by the effective radius\footnote{circular radius that encloses 50\%\ of the total galaxy flux} ($R_{eff,50}$) in F160W in CANDELS photometric catalogs. After binning the rest-UV and \ha\ emission maps for all galaxies in our sample, we then take the median for each radius bin to compute the average rest-UV and \ha\ 1D light profiles. We opt for a median over a mean when stacking to ensure that the light profiles are not biased by the brightest objects in the sample.

Figure~\ref{fig:result_reff_extended} (top panel) shows the average light profiles for our full sample along with the total number of galaxies and pixels in each bin of radius shown in the inset. All pixels that fall within the galaxies' segmentation map are considered when taking the median regardless of the per-pixel $SNR$. This allows us to naturally recover the noise properties of our data by measuring the error on the median (shown as the shaded confidence regions in Figure~\ref{fig:result_all_extended}) and we exclude any bins where the light profile is not well constrained (i.e., where both rest-UV and \ha\ stacked average light profiles are measured at $>1\sigma$ and at least $>$100 galaxies are contributing to stack in the corresponding bin). We do not apply any normalization to the individual galaxies' light profiles when stacking since we want to compute the \uvha\ flux ratio, $\zeta$, and need to preserve the inherent variations in the total flux.

\begin{figure}[!b]
\centering
\includegraphics[width=0.45\textwidth]{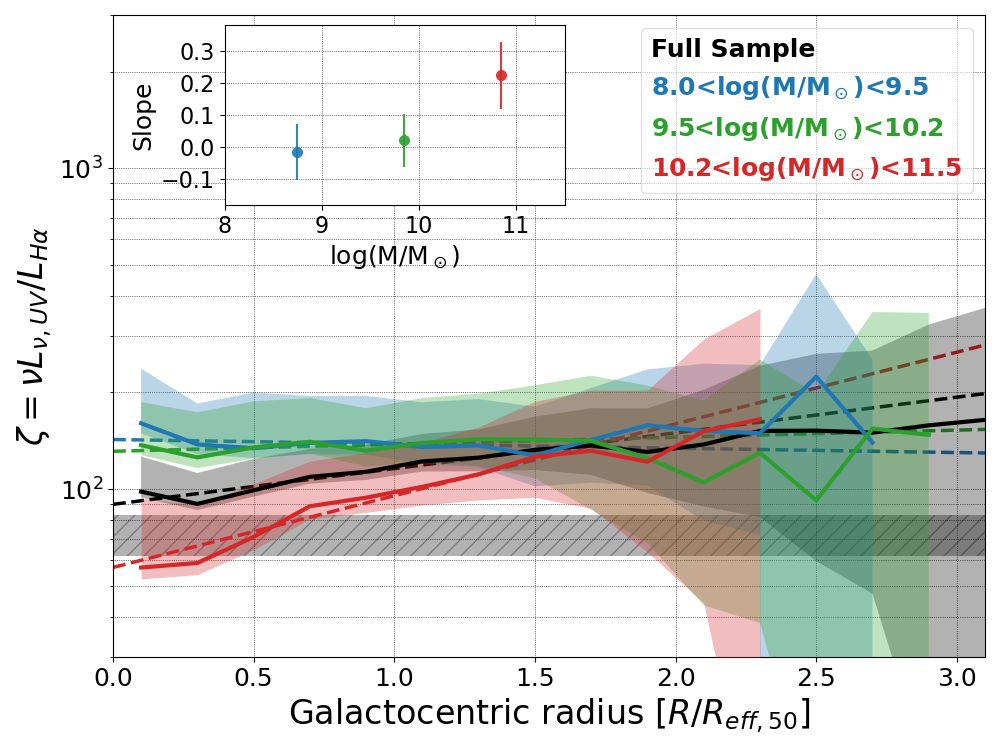}
\caption{The average \uvha\ ratio ($\zeta$) plotted as a function of the galactocentric radius for the full sample as well as three sub-samples split according to the galaxy stellar mass. The \textit{grey hatched, shaded} region shows the equilibrium value of $\zeta$ expected for constant star-formation. The \textit{shaded} bands denote the 1$\sigma$ errorbars. The \textit{dashed} lines are the best-fit linear slopes fitted at $R/R_{eff}<1.8$. The \textit{inset} shows the best-fit values of the slopes along with its $1\sigma$ uncertainty as a function of the mean galaxy stellar mass of the sub-sample.}
\label{fig:result_reff}
\end{figure}

\subsection{Average $\zeta$ as a function of galactocentric radius}
\label{sec:result_reff}

We compute the average burstiness parameter, $\zeta$, from the rest-UV and \ha\ light profiles as a function of galactocentric radius (shown in the bottom panel of Figure~\ref{fig:result_reff_extended}. When computing the average $\zeta$, we only consider bins where the measured median values of both rest-UV and \ha\ light profiles are significant ($>1\sigma$). As evident from the figure, $\zeta$ shows a clear trend as a function of galactocentric radius. The average $\zeta$ near the galactic centers ($R/R_{eff}<0.5$) is broadly consistent with the equilibrium value expected from constant star-formation. On the other hand, as we move towards the outskirts ($R/R_{eff}>1$), the average $\zeta$ trends upward deviating from the equilibrium value.

The stellar mass of a galaxy is one of the key physical quantities that is linked to its overall star-formation activity \citep[e.g.,][]{speagle14}. Hence, we extend our analysis to investigate the radial trend of $\zeta$ as a function of the galaxy stellar mass. Owing to the sensitivity and area of UVCANDELS and 3D-HST$+$AGHAST, our sample of 979 galaxies is large enough to allow us to split it into sub-samples of mass and still have sufficient number statistics for our analysis. We use the stellar mass estimated from SED fitting (Section~\ref{sec:sample}) to split our sample into three bins: low-mass ($<10^{9.5}$~M$_\odot$; 372~galaxies), intermediate-mass ($10^{9.5-10.2}$~M$_\odot$; 313~galaxies), and high-mass ($>10^{10.2}$~M$_\odot$; 294~galaxies).

Figure~\ref{fig:result_reff} shows the averaged $\zeta$ as a function of galactocentric radius now split according to three mass bins. There is clear and significant trend in the radius-dependent slope of $\zeta$ as a function of the stellar mass. We fit a linear slope to the inner region ($R/R_{eff}<1.8$) where the trend in $\zeta$ is most well-constrained. We find a best-fit slope of $-0.01\pm0.09$, $0.02\pm0.08$, and $0.22\pm0.10$ for the low-, intermediate- and high-mass bins, respectively, (also shown in the inset of Figure~\ref{fig:result_reff}) and $0.11\pm0.05$ for the full sample. We find a dispersion in $\zeta$ of 0.1$-$0.8~dex per bin in galactocentric radius (0.2~$R_{eff,50}$), with an increasingly larger dispersion towards larger radii. Given our light profiles (and $\zeta$ trends) are computed out to $R_{eff,50}$ of $\sim$3, these dispersion values are comparable to those reported in recent literature for integrated galaxy fluxes (e.g., 0.06$-$0.16~dex from \citealt{broussard22}). We discuss the implications of this result for the galaxy SFHs in Sections~\ref{sec:interpretation1} and \ref{sec:interpretation2}.

\begin{figure}[!t]
\centering
\includegraphics[width=0.45\textwidth]{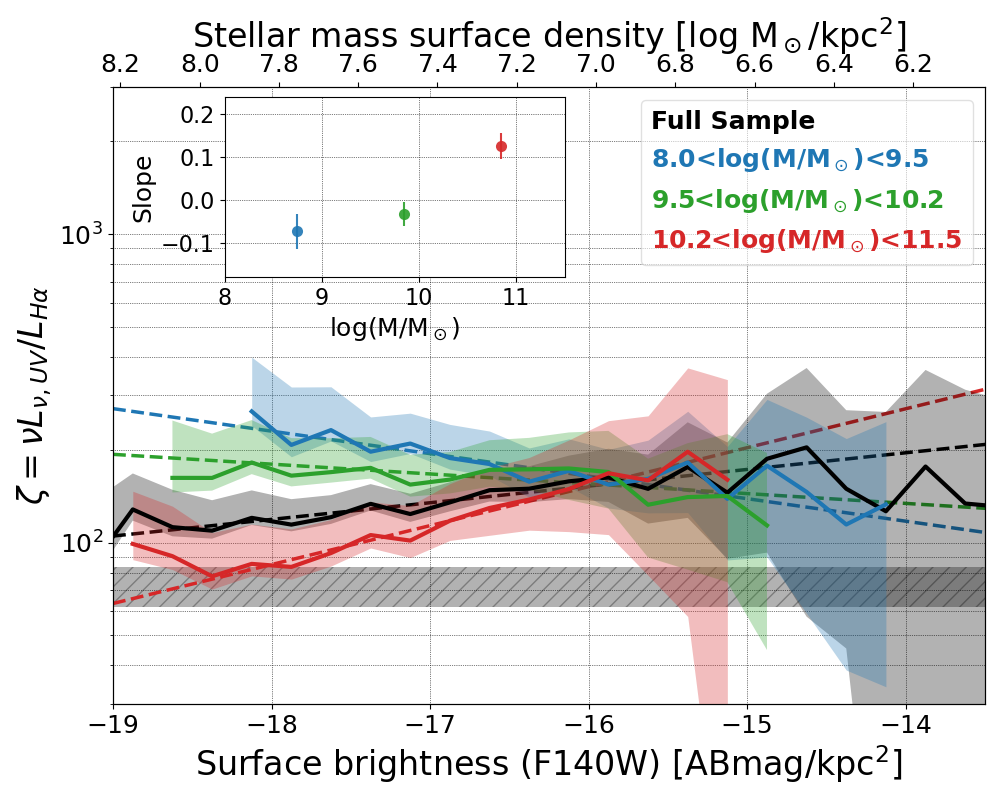}
\caption{Similar to Figure~\ref{fig:result_reff} but now as a function of the surface brightness in F140W (corrected for surface brightness dimming), which serves as a proxy for the mass surface density (shown as a twinned axis on top). The linear slopes are fit over the F140W surface brightness range of $[-18.5,-15.5]$.}
\label{fig:result_f140w}
\end{figure}

\subsection{Average $\zeta$ as a function of surface brightness}
\label{sec:result_f140w}

It is well established that the local stellar mass content (i.e. the mass surface density of existing stellar populations) is a key factor for regulating star-formation on galaxy scales \citep[e.g.,][]{kennicutt98,kennicutt12} and consequently, it is likely that the local SFH within the galaxy is also impacted by the stellar mass surface density.

While the mass surface density of galaxies generally falls off as a function of radius, the morphology of star-forming galaxies can often be irregular, especially at the redshifts of our sample \citep[e.g.,][]{scarlata07,patel13,shibuya16} and particularly at wavelengths that trace on-going star-formation activity. The azimuthally averaged $\zeta$ as a function of radius (as in Section~\ref{sec:result_reff}) marginalizes over key morphological features with active star-formation, e.g., star-forming clumps or spiral arms. Instead of the galactocentric radius, the local stellar mass surface density is a more direct tracer of the local environment in the galaxy at the sites of star-formation activity. Ideally, computing the stellar mass surface density would require a full-scale, multi-wavelength, resolved SED fitting, which is beyond the scope of this work. Instead, we utilize the surface brightness as a proxy that directly traces the stellar mass surface density.

We use the calibrated F140W stamps (the same as the ones used for morphological measurements in Section~\ref{sec:stacking}) for our sample after correcting for surface brightness dimming. The final stamps used for our analysis are units of ABmag~kpc$^{-2}$. The F140W filter traces optical wavelengths ($5500-8300$\angstrom) at the redshifts of our sample and thus is minimally sensitive to the on-going star-formation and a good tracer of the existing stellar mass. As demonstrated in \citep{dai21}, the $H$-band magnitude and galaxy stellar mass are strongly correlated for star-forming galaxies over the redshift range of our sample. Using the empirical relation from \citet{dai21}\footnote{We apply an average $k$-correction for our sample when converting surface brightness to stellar mass surface density when using the \citet{dai21} $H$-band absolute magnitude ($k$-corrected) to stellar mass relation.}, we can further interpret the F140W surface brightness as a stellar mass surface density.

We modify the analysis from Section~\ref{sec:stacking} binning the pixels according to the F140W surface brightness. Figure~\ref{fig:result_f140w} shows average $\zeta$ computed as a function of the F140W surface brightness for the full sample along with the three mass sub-samples described in Section~\ref{sec:result_reff}. We again fit linear slopes to the trend in $\zeta$ and find best-fit slope values of $-0.07\pm0.04$, $-0.03\pm0.03$, and $0.13\pm0.03$ for the low-, intermediate- and high-mass bins, respectively, and $0.05\pm0.02$ for the full sample. All the slopes are much better constrained and their evolution as a function of the galaxy stellar mass is much more pronounced than the case with radius. The implications of these trends are discussed in Section~\ref{sec:interpretation3}.

\begin{figure*}[!t]
\centering
\includegraphics[width=0.8\textwidth]{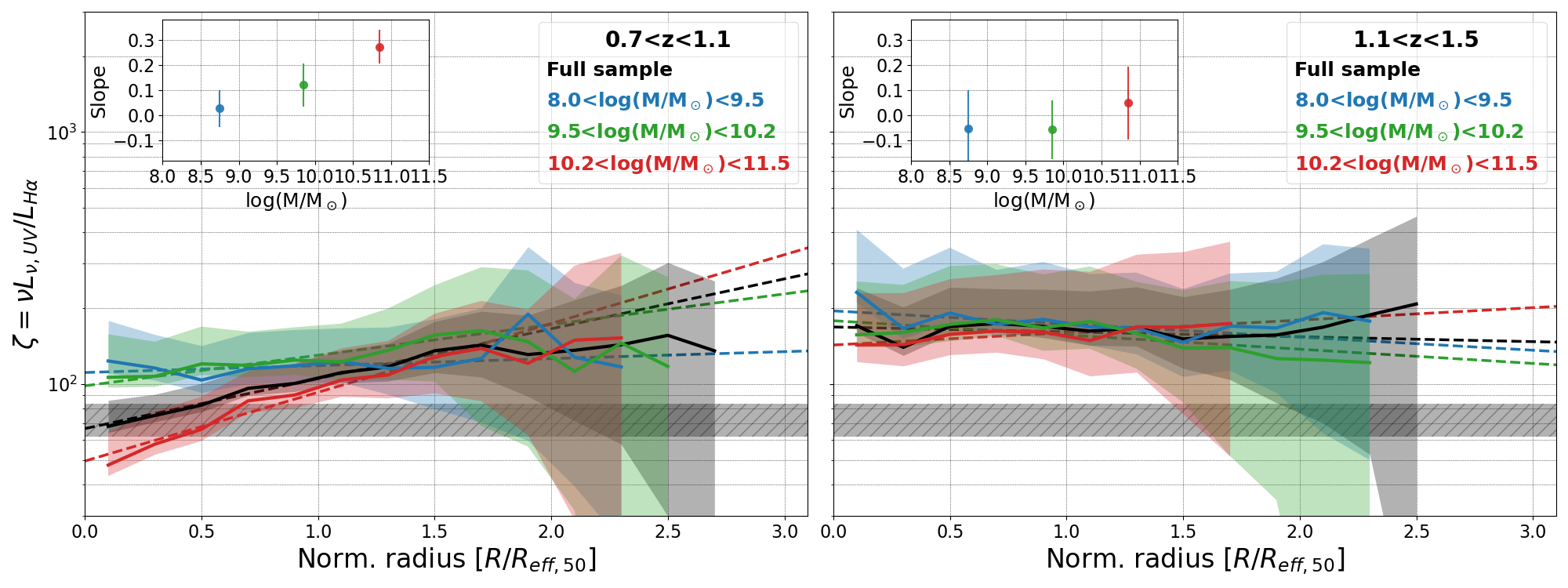}\\
\includegraphics[width=0.8\textwidth]{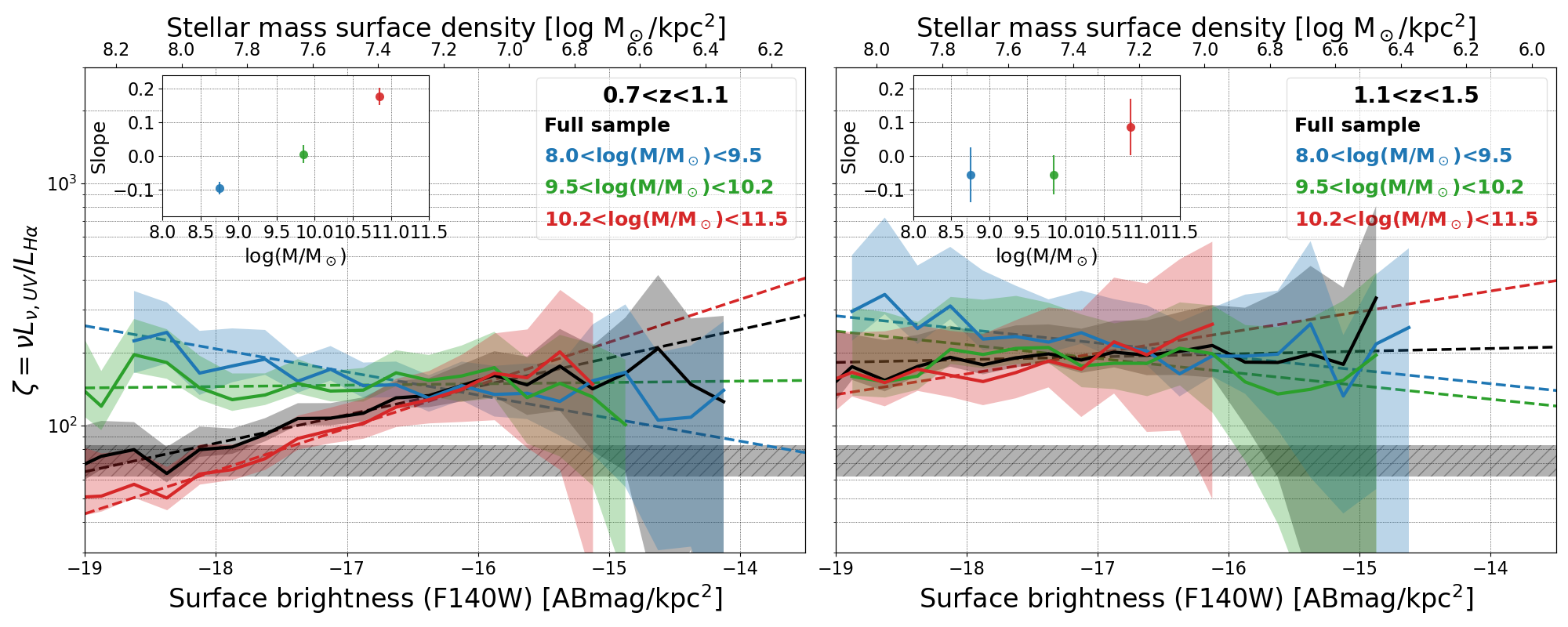}
\caption{Similar to Figures~\ref{fig:result_reff} and \ref{fig:result_f140w} but now split into two redshift bins: $0.7<z<1.1$ and $1.1<z<1.5$.}
\label{fig:result_zbins}
\end{figure*}

\subsection{Average $\zeta$ trends over redshift}
\label{sec:result_zbins}

The redshift range around $z \sim 1$ is at the tail end of the \textit{cosmic high noon} and is a time where the cosmic star-formation density is evolving rapidly. Given the large galaxy sample size considered in this work, we are able to further split our analysis into bins of redshift to investigate any potential evolution in the $\zeta$ trends as a function of cosmic time.  In order to ensure sufficient number statistics for all the stellar mass sub-bins, we split our sample at $z=1.1$, into two redshift bins: $0.7<z<1.1$ and $1.1<z<1.5$ with sample sizes of 508 and 471 galaxies, respectively.

Figure~\ref{fig:result_zbins} shows the main result from our analysis, i.e., the trend in $\zeta$ as a function of galactocentric radius and F140W surface brightness, for the two bins in redshift along with the three mass sub-samples described in Section~\ref{sec:result_reff}. Overall, the $\zeta$ trends are consistent with flat slopes for the higher redshift ($1.1<z<1.5$) bin, both as a function of radius as well as surface brightness. On the other hand, in the lower redshift ($0.7<z<1.1$) bin, the slope of the $\zeta$ trends get progressively steeper as a function of the galactocentric radius. In the surface brightness case, the lower redshift sample exhibits strong evolution in $\zeta$ for the low- and high-mass bins with significantly negative and positive slopes for each case, respectively. The slope values for all sub-samples in mass as well as redshift are reported in Table~\ref{tab:slope_values} and we further discuss the $\zeta$ evolution with redshift in Section~\ref{sec:interpretation4}.

\section{Discussion}
\label{sec:discussion}

\subsection{Impact of key assumptions}
\label{sec:caveats}
Before delving into discussing the implications of our results from Section~\ref{sec:results} in the context of galaxies' star-formation properties, it is important to consider the impact of various assumptions made during the analysis.

\textbf{Dust correction:}
Our fiducial case uses the resolved dust maps (described in Section~\ref{sec:maps_dust}) along with the \citet{calzetti00} dust law. To assess the impact of these assumptions, we test two additional scenarios. First, we switch to using the SMC \citep{gordon03} dust law along with the appropriate calibration from \citet{reddy18}. The steeper SMC dust law causes the dust-corrected light profiles to be fainter overall (shown in dashed curves in left panel of the Figure~\ref{fig:result_reff_extended}). We also check for a dust law from \citet{buat18} that is directly calibrated for $z \sim 0.6-1.6$ galaxies using \textit{HST} grism spectroscopy (shown as dotted curves in the left panel of Figure~\ref{fig:result_reff_extended}). In both these cases, the impact of changing the dust law on $\zeta$ is minimal as it only affects the overall normalization of the trend in $\zeta$ and does not significantly affect its shape.

Secondly, we evaluate the impact of using resolved dust maps (from Section~\ref{sec:maps_dust}) when correcting for dust attenuation. We recompute the light profiles assuming a single value of dust for the whole galaxy, instead of resolved $A_V$ maps. This has a minimal impact on the overall result (shown as dotted curves in Figure~\ref{fig:result_reff_extended} and \ref{fig:result_all_extended}). The differences from the fiducial case are most evident in the inner regions which is to be expected since that is where the individual pixels would have sufficient $SNR$ to facilitate a reliable per-pixel measurement of the UV slope $\beta$. Importantly, even in the inner regions, the overall light profiles as well as the trend in $\zeta$ are minimally impacted (see also Appendix~\ref{appndx:results_all} for additional figures).

Similarly, we also perform the converse test evaluating the impact on the stacked results if we only consider the pixels where reliable per-pixel dust measurement is possible. The impact of this test on the trends in $\zeta$ is even more minimal (shown as dashed-dotted curves in Figures~\ref{fig:result_reff_extended} and \ref{fig:result_all_extended2}). The resulting trends from this case are nearly identical to the fiducial case, with the primary difference being that the light profiles have more noise, particularly in the outskirts due to there being fewer pixels with high enough $SNR$ for a reliable per-pixel dust estimate. From these tests, we are confident that the observed trend in $\zeta$ is not driven by systematic biases from our particular methodology for dust correction. Overall, our results are robust against our choice of using resolved dust maps as well as our dust law assumption.

Lastly, the factor relating stellar and nebular dust attenuation,  $f=E(B-V)_{stellar}/E(B-V)_{nebular}$ also affects the computed $\zeta$ trends. For our fiducial result, we have assumed a value of 0.44; however, larger values of $f$ have been suggested from observational studies up to $f \sim 1$ \citep[e.g.,][]{puglisi16,rodriguez22}. The primary impact of changing $f$ to larger values is that the trend in $\zeta$ is shifted to higher values, but most importantly, the slope of the $\zeta$ relation is not affected significantly, thus having a minimal impact on our conclusions.

\textbf{\nii\ Correction:}
In this analysis, we apply the mass- and redshift-dependent correction from \citet{faisst18} when correcting the blended \ha+\nii\ flux from the grism observations. While this accounts for the evolution in the \nii/\ha\ ratio over cosmic time as well as its variation due to global galaxy properties, we still assume a single correction value for the whole galaxy, where as galaxies are known to have metallicity gradients, where the outskirts of galaxies tend to have lower metallicities than their cores \citep[e.g.,][]{jones15,wang20,wang22,simons21,li22}. A metallicity gradient could result in a gradient in the \nii/\ha\ ratio within a galaxy, which in turn would affect $\zeta$. A lower \nii/\ha\ ratio would require a smaller correction and thus result in a lower value of $\zeta$. However, the metallicity gradients are expected to typically be on the order of $\lesssim0.2$~dex~kpc$^{-1}$ even at high redshifts \citep[e.g.,][]{wang20,wang22}. The change in $\zeta$ due to such a gradient is negligible relative to the observed trends and is well within the 1$\sigma$ error bars of our results. While these gradients may play a role in driving the trends in $\zeta$, it is not the dominant factor and the trends in our results are not exclusively driven by the presence of metallicity gradients in galaxies.

\textbf{IGIMF:}
In Section~\ref{sec:burstiness_parameter}, we consider the impact of various canonical IMFs that have traditionally been universally applied to galaxies. On the other hand, a non-universal IMF, such as the integrated galactic initial mass function \citep[IGIMF,][]{weidner05,pflamm07,pflamm09,weidner11} predicts an IMF slope that gets steeper than the canonical case with decreasing total SFR of galaxies. The IGIMF predicts a lower chance of forming massive stars in low-SFR galaxies compared to their higher SFR counterparts. Due to the lack of massive stars, an IGIMF results in low-SFR galaxies having a higher value of $\zeta$. \citet{pflamm09} demonstrate the impact of IGIMF on $\zeta$ and find that for galaxy SFRs $<10^{-2}$~M$_\odot$~yr$^{-1}$, there is a sharp up-turn in $\zeta$ under the IGIMF scenario.

However, the SFR values where this effect is most significant are lower than those considered in our analysis. None of the galaxies in our sample have SFRs~$<10^{-2}$~M$_\odot$~yr$^{-1}$, with the majority having SFRs $\gtrsim 10^{-1}$~M$_\odot$~yr$^{-1}$. Even for our fiducial result of resolved $\zeta$, the rest-UV and \ha\ light profiles are limited to a surface brightness of $\nu L_\nu \gtrsim 10^{41.8}$~\ergskpc\ and $L_{H\alpha} \gtrsim 10^{39.6}$~\ergskpc, corresponding to a SFR of $\gtrsim 10^{-1.7}$~M$_\odot$~yr$^{-1}$~kpc$^{-2}$ according to the standard \citep{kennicutt12} calibration. Considering the total area in each of our bins (0.2~$R_{eff,50}$) and the range of $R_{eff,50}$ for our sample (1$-$10.6~kpc with a median of $\sim$2.5~kpc), the total SFR is $\gtrsim$10$^0$~M$_\odot$~yr$^{-1}$ for all bins analyzed in this work, which is considerably higher than the regime where the effects of an IGIMF are significant \citep{pflamm09}.

\textbf{Using surface brightness as a proxy for mass surface density:}
The young stars in regions with active star-formation will have a non-trivial contribution at all wavelengths, including the observed $H$-band (rest-frame optical). In the case where the relative contribution at rest-frame optical wavelengths from the young stars is comparable to that from the existing stellar population, the F140W flux may be boosted due the ongoing star-formation. This effect would be most significant in the less dense regions of the galaxy, where it would result in a increased F140W surface brightness relative to the fiducial assumption of it tracing just the existing stellar population. While this is a second order effect, this scenario is plausible in the low-mass galaxies or galaxy outskirts, in general, and could affect our results from stacking as a function of F140W surface brightness (Section~\ref{sec:result_f140w}).

On the other hand, stacking as a function of galactocentric radius (Section~\ref{sec:result_reff}) is immune to this effect and hence, comparing the results from Figures~\ref{fig:result_reff} and \ref{fig:result_f140w}, we can confirm that the impact of this effect is negligible for the intermediate- and high-mass bins. This effect could be the cause for the slight negative $\zeta$ slope for the low-mass sample as a function of the F140W surface brightness (Figure~\ref{fig:result_f140w}). However, since we already confirmed a flat slope as a function of galactocentric radius for the low-mass sub-sample in Figure~\ref{fig:result_reff}, we do not anticipate this effect having any further impact other than potentially flattening the $\zeta$ slope as a function of F140W surface brightness, making it consistent with our result in Section~\ref{sec:result_reff}.

\subsection{Implications on burstiness in galaxies' SFH}
\label{sec:interpretation}

\subsubsection{Radial evolution of $\zeta$}
\label{sec:interpretation1}
As discussed in Section~\ref{sec:burstiness_parameter}, the galaxy SFH is the one of the dominant factors that drives the deviations in the value of $\zeta$ relative to its equilibrium value from the constant star-formation case. The presence of bursts causes $\zeta$ to scatter asymmetrically about the equilibrium value, where it deviates to preferentially larger values following a burst. In this analysis, we have computed the average $\zeta$ as a function of galactocentric radius (Section~\ref{sec:result_reff}) and rest-frame optical surface brightness (Section~\ref{sec:result_f140w}) for our galaxy sample via a stacking analysis, providing an overall trend in $\zeta$ for a statistically significant sample of galaxies.

In Figure~\ref{fig:result_reff_extended}, we find that the average $\zeta$ from stacking our full galaxy sample ($0.7<z<1.5$; $10^{8-11.5}$~$M_\odot$) has a significant positive slope as a function of the galactocentric distance, where $\zeta$ trends upwards toward larger distances. Given that this value of $\zeta$ is a statistic averaged across a population of galaxies, it indicates that a substantial fraction of galaxies (or pixels) have $\zeta$ that is higher than the equilibrium value and suggestive of the presence of bursts in the galaxies' SFHs. At face value, the observed trend in $\zeta$ suggests that, on average, the outskirts of the galaxies in our sample are experiencing an increased contribution from star-bursts relative to their cores.

It is also important to point out that bursts of star-formation are not the only explanation for deviations in $\zeta$. As discussed in Section~\ref{sec:burstiness_parameter}, long-term smoothly declining (rising) SFHs can also result in a value of $\zeta$ that is higher (lower) than the equilibrium value. The higher-than-equilibrium $\zeta$ values from Figure~\ref{fig:result_reff_extended} could also be indicative of a smoothly declining (or recently declined and still low) SFR. In this case, the observed trend would suggest a sharper decline in the galaxies' SFHs in their outskirts relative to their cores. Given the degeneracy in the behavior of $\zeta$ in these two scenarios, we are unable to distinguish between them within the context of this analysis and thus long-term smoothly rising/falling SFHs remain as a plausible explanation for the trends in $\zeta$ presented in this work.

\subsubsection{Impact of galaxy mass on the radial evolution of $\zeta$}
\label{sec:interpretation2}
In Figure~\ref{fig:result_reff}, we present the average $\zeta$ trends as a function of galactocentric distance for three sub-samples split according to the galaxy stellar mass to investigate the impact of the galaxy mass on the trends in $\zeta$ (see also Appendix~\ref{appndx:results_all} for the average light profiles and $\zeta$ profiles for the individual mass bins). For the low- ($<10^{9.5}$~M$_\odot$) and intermediate-mass ($10^{9.5-10.2}$~M$_\odot$) bins, the average $\zeta$ is roughly constant at all radii, albeit at a level that is elevated above the equilibrium value from constant star-formation. This suggests that the SFHs in these galaxies do not change significantly as a function of radius. The higher-than-equilibrium value of $\zeta$ also potentially indicates that these galaxies have a significant contribution from bursts at all radii. This is consistent with studies that use the integrated galaxy flux; e.g., \citet{guo16} who find, for $0.4<z<1$, galaxies below 10$^9$~M$_\odot$ to be dominated by bursty SFHs.

On the other hand, the high-mass bin has a steep slope in $\zeta$ as function of radius, where the interior regions ($R/R_{eff}<1$) are consistent with the equilibrium value, but their outskirts ($R/R_{eff}>1.5$) clearly appear to have a bursty component to their SFHs. The strong evolution in $\zeta$ for the high-mass bin is largely responsible for driving the overall trend for the full sample in Figure~\ref{fig:result_reff_extended}.

Existing studies of high-mass galaxies typically do not find significant evidence of burstiness in their SFHs \citep[e.g.,][]{lee09,weisz12,guo16,emami19}; however, these findings are based on the full integrated photometry. What we uncover with this work, instead, is that even for these massive galaxies, the star-formation occurring in their outskirts has a significant burstiness. It is only when taken as a whole that the brighter central regions dominate the total flux contribution and cause the overall SFH of the galaxy to appear consistent with secular star-formation with no significant bursts.

\subsubsection{Evolution of $\zeta$ as a function of surface brightness}
\label{sec:interpretation3}
We find the trends in $\zeta$ (shown in Figure~\ref{fig:result_f140w}) to be more tightly correlated with the rest-frame optical surface brightness, which traces the galaxies' morphological features compared to simply the galactocentric distance. In high-mass galaxies, on average for the full sample, the observed $\zeta$ in brighter (and thus denser) regions ($\lesssim-17$~ABmag~kpc$^{-2}$ or $\gtrsim10^{8.4}$~M$_\odot$~kpc$^{-2}$) of the galaxy is consistent with the equilibrium value from the constant star-formation scenario. This average trend over the full sample is driven largely by the lower redshift ($z<1.1$) galaxies (see Section~\ref{sec:interpretation4} for further discussion). On the other hand, for regions that are fainter (less dense), the average $\zeta$ value suggests a significant contribution from bursts to the SFH. This is similar to the trend from Figure~\ref{fig:result_reff} which is expected considering that galactocentric radius is a good tracer of the mass surface density, especially for high-mass galaxies that tend to have smoother morphologies.

On the other hand, for low-mass galaxies, we find an opposite trend such that the burstiness in the SFHs is intensified toward brighter (higher density) regions. It is likely that for these low-mass and often compact galaxies, the bursts occurring in the brighter (denser) regions are more intense. The strength of individual star-bursts relative to the underlying, low-level constant SFR affects the amplitude of the deviation in $\zeta$ in the post-burst phase. This could present a plausible explanation for the observed trend, where bursts in the denser regions of low-mass galaxies are more intense and the resulting stellar feedback could more easily eject the existing cold gas and/or prevent/delay accretion in the neighboring less dense regions. However, as discussed in Section~\ref{sec:caveats}, for the low-mass sample, the on-going star-formation could be contributing significantly to the rest-frame optical surface brightness, thus rendering it ineffective as a tracer of the existing stellar population. This effect could also be a potential reason for the slight negative slope in $\zeta$ as a function of surface brightness. Lastly, we reconfirm our findings from the radial case that, on average, low- and intermediate-mass bins have significant contributions from bursts in their SFHs, regardless of the local surface brightness.

Most importantly, our analysis of $\zeta$ as a function of the surface brightness leads us toward a seemingly fundamental observation. From the trends shown in Figure~\ref{fig:result_f140w}, we note that for any given galaxy, regardless of its stellar mass, local regions with a mass surface density below $\sim$10$^{7.5}$~M$_\odot$~kpc$^{-2}$ appear to have a significant bursty component in their SFHs. This could be an indication that, at the smallest scales, the local SFH is independent of the global galaxy properties. This delimiting value in mass surface density is provided as a rough estimate and we caution against treating it as exact, since it is based on an empirical relation between the $H$-band flux and stellar mass from \citet{dai21}, which is defined for integrated galaxy fluxes/stellar masses, has considerable scatter, and is not as tightly constrained down to $\lesssim$10$^8$~M$_\odot$.

\subsubsection{Evolution of $\zeta$ trends with redshift}
\label{sec:interpretation4}

As per Section~\ref{sec:result_zbins}, we split our analysis into two redshift bins: $0.7<z<1.1$ and $1.1<z<1.5$, in order to investigate the evolution in $\zeta$ over this critical time period when the universe is undergoing a rapid evolution in its cosmic SFR density. As shown in Figure~\ref{fig:result_zbins}, we find a clear distinction in the $\zeta$ trends for $z>1.1$ galaxies. The $\zeta$ trends as a function of both galactocentric radius as well as surface brightness are consistent with being flat for $z>1.1$ galaxies. At face value, this seems to suggest that the star-formation properties of the general star-forming population at $z>1.1$ are similar across the full range of stellar masses from 10$^8$ to 10$^{11.5}$~M$_\odot$ with them appearing to have a consistently bursty SFH with seemingly no dependence on radius or surface brightness.

On the other hand, the $z<1.1$ galaxy population appears to significantly deviate from flat $\zeta$ trend suggesting that there is indeed an evolution in the star-formation properties within individual galaxies toward the end of the \textit{cosmic high-noon}. As the average cosmic star-formation activity is declining, the central (denser) regions of galaxies appear to settle into a more secular star-formation activity, whereas the outskirt (less dense) regions continue to experience bursts in their SFHs as they did during the peak of \textit{cosmic high-noon}. Lastly, splitting the analysis in redshift also reveals that the $\zeta$ trends observed for the full galaxy sample (see the discussions in Sections~\ref{sec:interpretation1} and \ref{sec:interpretation2}) are primarily driven by the low-redshift ($z<1.1$) galaxies in our sample.

\subsubsection{Potential link between burstiness and stellar clumps}
Star-forming clumps observed in galaxies may present an interesting connection to SFH burstiness explored in this work. Giant, star-forming clumps are known to be prevalent in galaxies with stellar masses as high as $10^9$~M$_\odot$ and specific SFRs up to $10^{-7}$~yr$^{-1}$. While they may be rare in the local universe \citep{fisher17,mehta21,adams22}, these clumps are found in abundance at high-redshifts \citep[$0.5<z<2$; e.g.,][]{guo15,guo18,shibuya16} with the fraction of star-forming galaxies containing clumps reaching $\sim60\%$ at $z\sim2$ \citep{guo15}. These clumps are sites of active star-formation and it is possible that these are linked to the burstiness in galaxy SFHs. Clumps have been confirmed to have a higher SFR compared to their galactic surroundings and could be the physical manifestations of the bursts in galaxy SFHs.

\citet{guo15} find a significantly higher clumpy fraction for low-mass galaxies ($<10^{9.8}$~M$_\odot$) relative to their high-mass counterparts. This has similarities to our result where low-mass galaxies are found to be more bursty, in general. The clumps found in outskirts of galaxies are found to have higher SFR than those in the inner regions \citep{guo18}, which is also consistent with our result which finds an increasing burstiness at larger galactocentric distances. While no explicit link has been established between burstiness in galaxy SFH and clumps thus far, it remains as an interesting avenue to explore both from an observational as well as a theoretical perspective.

\section{Conclusions}
\label{sec:conclusions}

The combination of UVCANDELS WFC3/UVIS imaging along with 3D-HST$+$AGHAST WFC3/IR slitless grism spectroscopy has provided a unique opportunity to study the resolved rest-frame UV and \ha\ emission from the same galaxies at $z\sim1$. Both rest-frame UV and \ha\ are excellent tracers of on-going star-formation activity in galaxies; however, since they are sensitive to different time-scales, a comparison of the two enables us to infer the burstiness in the SFH.

In this work, we select a sample of 979 star-forming galaxies over $0.7<z<1.5$ to compare their resolved rest-frame UV and \ha\ emission and investigate the burstiness of their SFHs as a function of their galaxy structural parameters. We start by homogenizing the rest-UV and \ha\ maps by aligning, calibrating, and PSF-matching as well as correcting them for dust. We then compute the \uvha\ ratio ($\zeta$) which serves as a means to parameterize the ``burstiness'' in star-formation. We perform a stacking analysis to generate averaged rest-UV and \ha\ light profiles and compute the evolution of $\zeta$ as a function of the galactocentric distance as well as the surface brightness. We perform extensive tests to ensure that the observed trends are not driven by any of the assumptions involved in our analysis, including dust corrections, IMF assumptions, and \nii\ correction applied to the blended \ha+\nii\ in grism observations.

Our main results are as follows:
\begin{itemize}
    
    \item We find the $\zeta$ averaged over the full galaxy sample to increase toward larger galactocentric distances. An increasing contribution from a bursty SFH toward the outskirts is one plausible explanation for this trend. A smoothly declining SFH with an increasingly sharper decline toward the galactic outskirts can also explain the observed trend.
    
    \item We further split our sample into three bins according to the galaxy stellar mass. We find the low- and intermediate-mass samples exhibit $\zeta$ that is roughly constant and indicative of little-to-no evolution in the galaxy SFH as a function of radius. The higher-than-equilibrium value of $\zeta$ is also indicative of bursty SFHs, which is consistent with existing studies that suggest galaxies below $\sim$10$^9$~M$_\odot$ to be dominated by bursty star-formation.
    
    \item On the other hand, we find a significant radial evolution in $\zeta$ for the high-mass sample, consistent with secular star-formation in the interior regions ($R/R_{eff}<1$) but rapidly deviating from the equilibrium value toward the outskirts ($R/R_{eff}>1.5$), potentially indicative of the outskirts having a bursty star-forming component. This suggests that, while the integrated photometry of high-mass galaxies does not indicate the presence of burstiness in their SFHs, the star-formation in their outskirts could still have a significant bursty component that is likely being overshadowed by the brighter cores that dominate the flux contribution.
    
    \item We also compute $\zeta$ as a function of the surface brightness (in F140W) which serves as a proxy for galaxy stellar mass surface density. We recover similar trends as in the case with radius, albeit they are more significant, suggesting that the local stellar mass surface density is the more fundamental factor in regulating the star-formation activity, rather than just distance from the center.
    
    \item Interestingly, we find that regardless of the galaxy stellar mass, regions with a mass surface density below $\sim$10$^{7.5}$~M$_\odot$~kpc$^{-2}$ have a significantly bursty star-formation, potentially indicating that local star-formation is independent of global galaxy properties at the smallest scales.

    \item Lastly, we find that the evolution of $\zeta$ as a function of the radius and surface brightness is most strongly driven by the star-forming population at $z<1.1$. The galaxies at $z>1.1$ appear to have flat $\zeta$ trends, indicating that their star-formation histories, while still bursty, do not evolve with radius or surface brightness.

\end{itemize}

The trends in $\zeta$ presented in this work will serve as key observational constraints for theoretical galaxy models. A comparison between these observational results and current state-of-the-art hydro-dynamical simulations that include the effects of stellar feedback and have the resolution to capture star-formation on $<10$~Myr timescales is bound to yield crucial insights into our understanding of how star-formation proceeds in galaxies. We leave this analysis for a future publication.

Future observations with JWST will make \ha\ accessible at higher redshifts ($z>1.5$) and will also push to fainter, lower mass galaxies than those considered in this work. Combining these with HST/UVIS imaging will allow us to push this analysis to earlier galaxies and uncover the properties of their star-formation processes at the peak of \textit{cosmic noon}. \\

\begin{acknowledgments}
We would like to acknowledge Dr. Gabriel Brammer for their critical contribution in the \grizli\ reduction of 3D-HST grism spectroscopy data as part of the Complete Hubble Archive for Galaxy Evolution (CHArGE) initiative. The analysis presented in this paper is based on observations with the NASA/ESA Hubble Space Telescope obtained at the Space Telescope Science Institute, which is operated by the Association of Universities for Research in Astronomy, Incorporated, under NASA contract NAS5-26555. Support for Program Number HST-GO-15647 was provided through a grant from the STScI under NASA contract NAS5-26555. V.M., H.T., and X.W. acknowledge work carried out, in part, by IPAC at the California Institute of Technology, as was sponsored by the National Aeronautics and Space Administration.
\end{acknowledgments} 

\facilities{HST (WFC3), HST (ACS)}

\software{NumPy, SciPy, AstroPy, Matplotlib, \grizli\ \citep{brammer21}, \swarp\ \citep{swarp}, \pypher\ \citep{pypher}, \dbasis\ \citep{dbasis}} 

\appendix

\section{Individual light profiles for sub-samples}
\label{appndx:results_all}

\begin{figure*}[!t]
\centering
\includegraphics[width=0.32\textwidth]{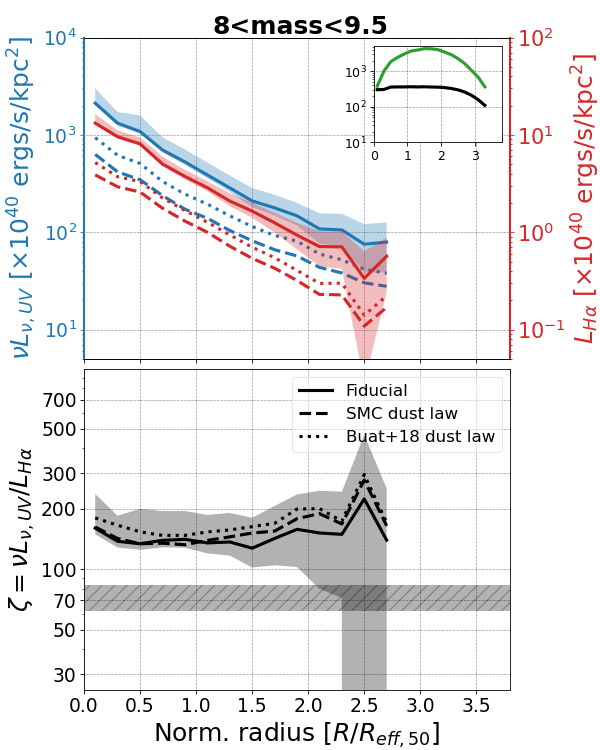}
\includegraphics[width=0.32\textwidth]{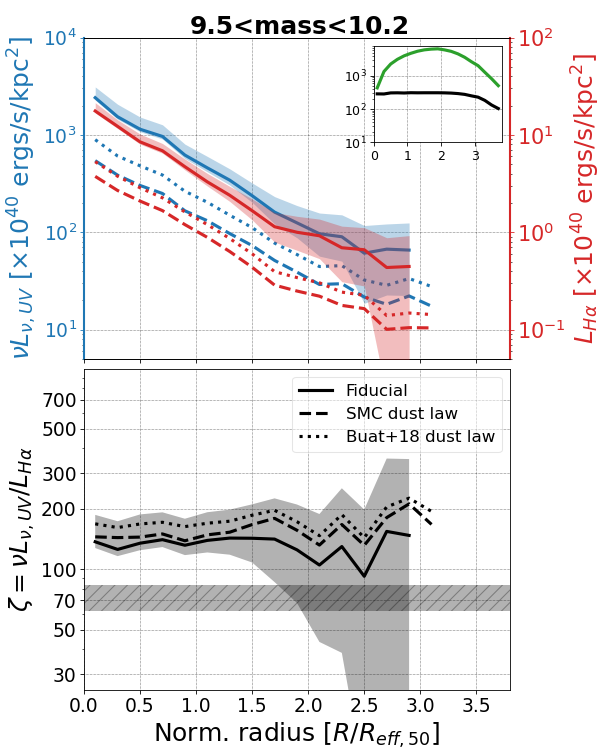}
\includegraphics[width=0.32\textwidth]{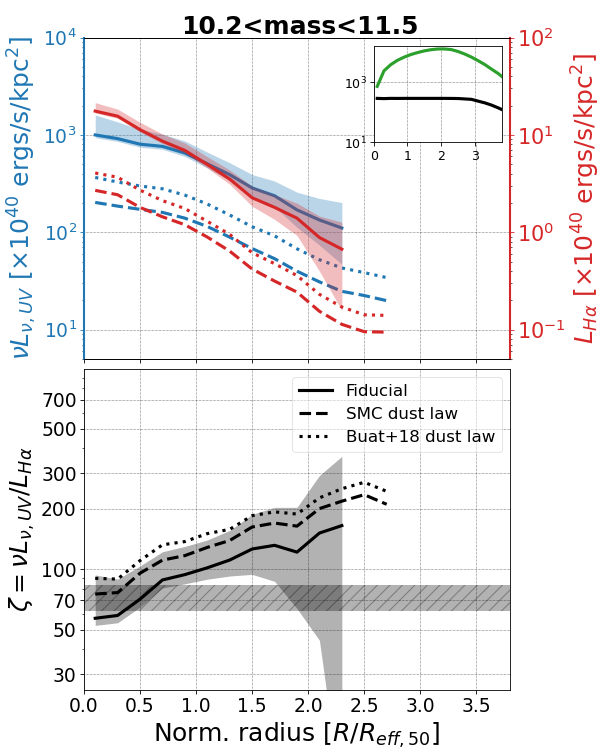} \\
\includegraphics[width=0.32\textwidth]{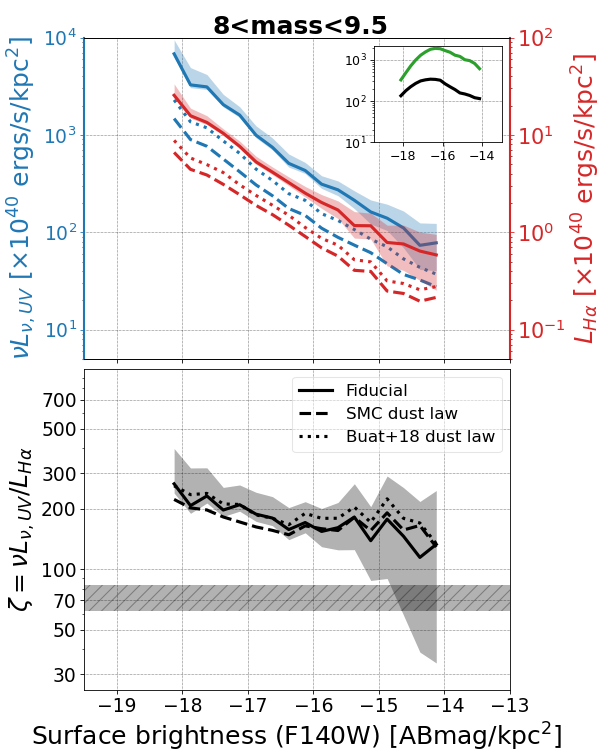}
\includegraphics[width=0.32\textwidth]{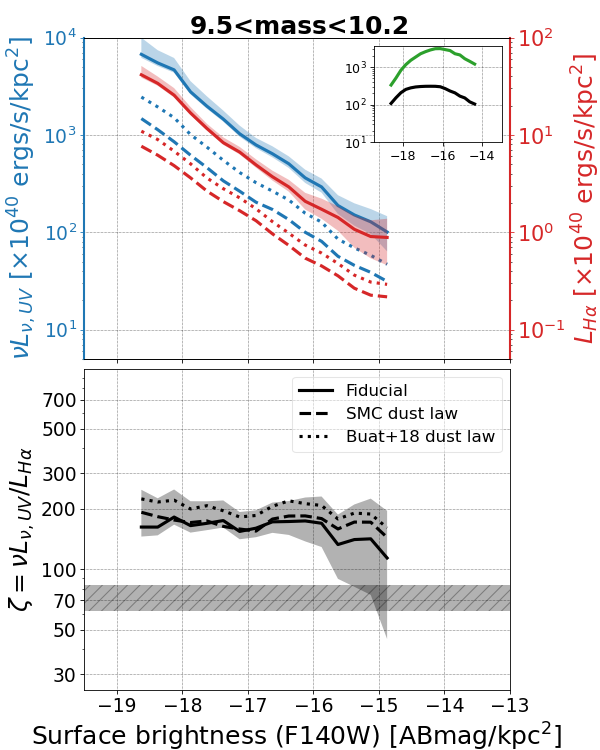}
\includegraphics[width=0.32\textwidth]{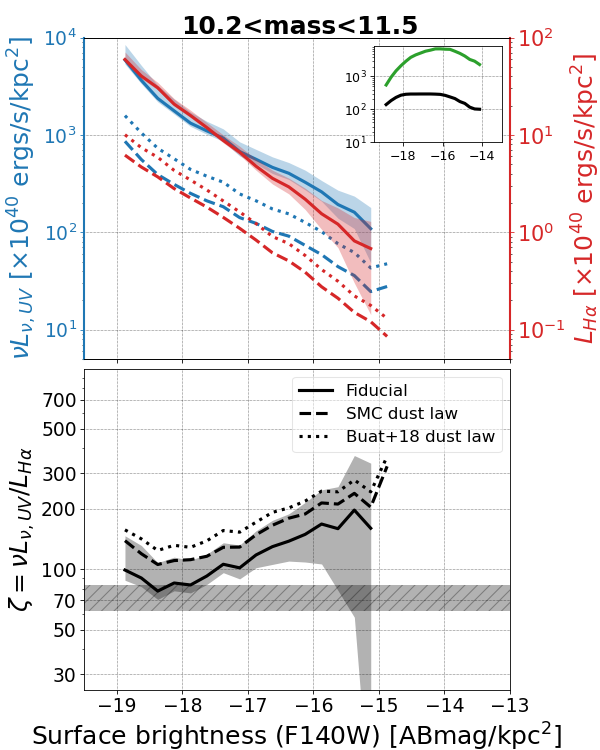}
\caption{Similar to the left panel of Figure~\ref{fig:result_reff_extended} but shown for each sub-sample binned in galaxy stellar mass and as a function of galactocentric distance (\textit{top} row) as well as surface brightness (\textit{bottom} row).}
\label{fig:result_all_extended}
\end{figure*}

\begin{figure*}[!t]
\centering
\includegraphics[width=0.32\textwidth]{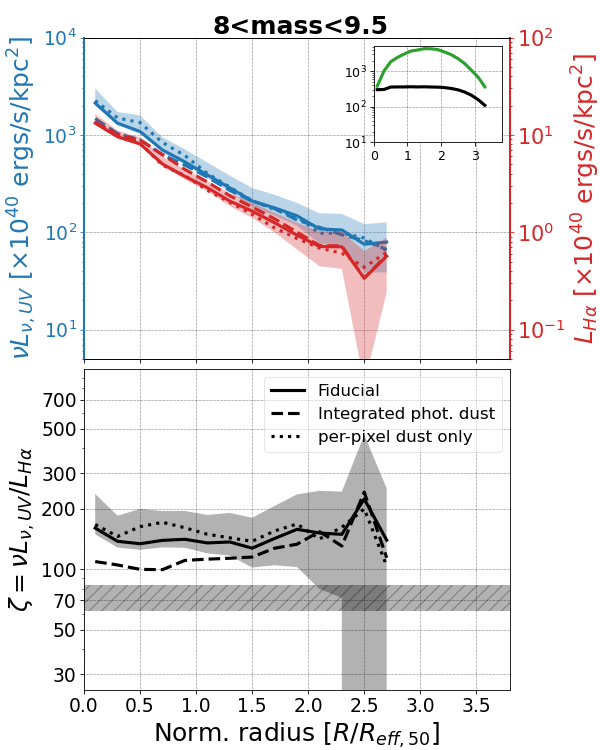}
\includegraphics[width=0.32\textwidth]{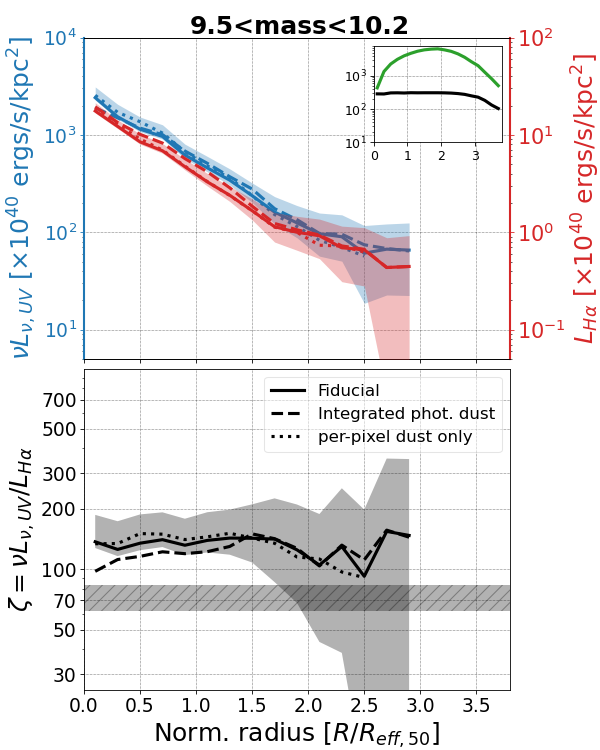}
\includegraphics[width=0.32\textwidth]{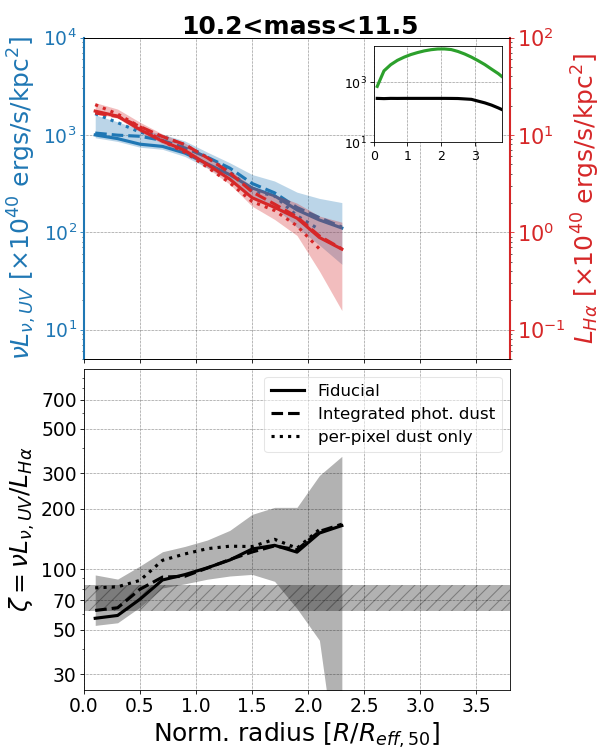} \\
\includegraphics[width=0.32\textwidth]{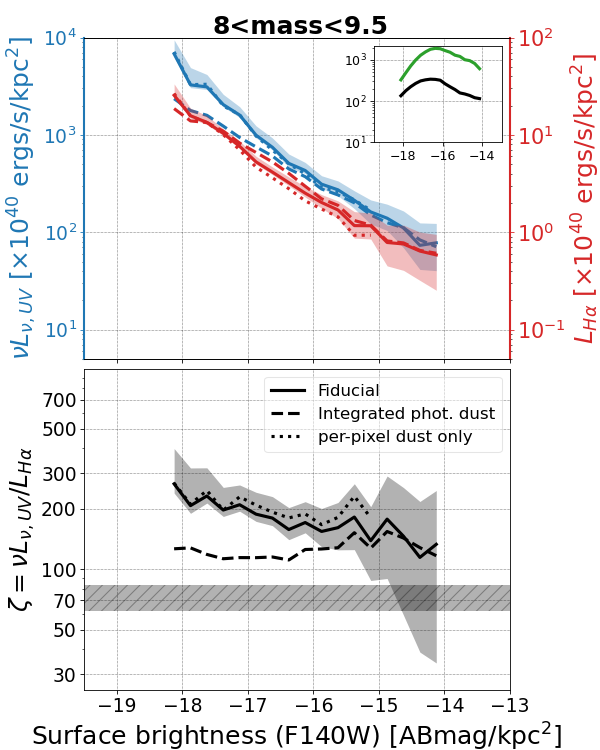}
\includegraphics[width=0.32\textwidth]{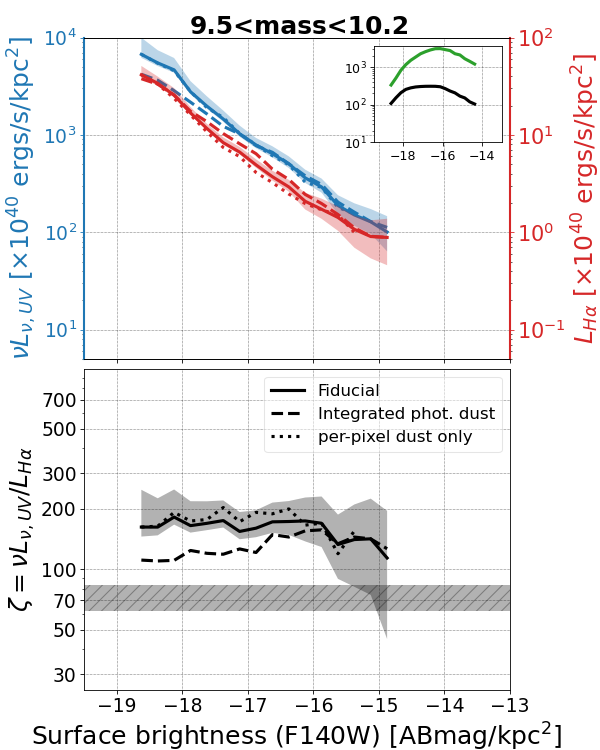}
\includegraphics[width=0.32\textwidth]{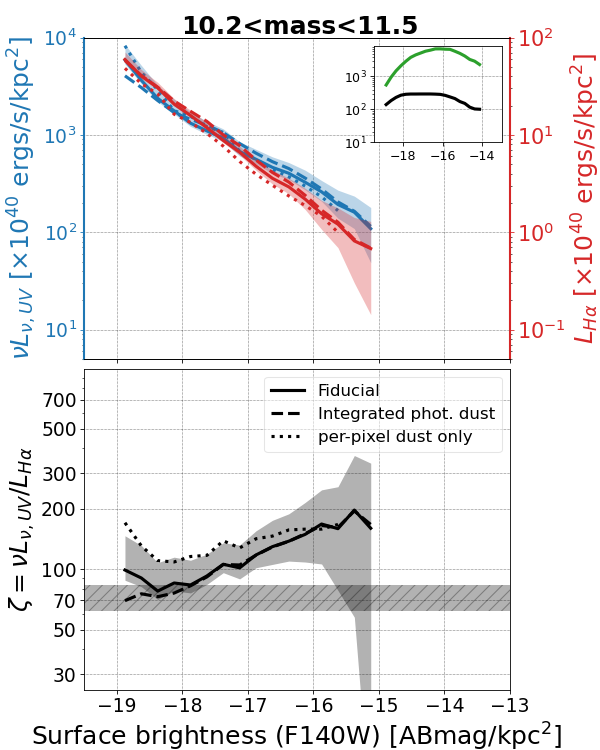}
\caption{Similar to the right panel of Figure~\ref{fig:result_reff_extended} but shown for each sub-sample binned in galaxy stellar mass and as a function of galactocentric distance (\textit{top} row) as well as surface brightness (\textit{bottom} row).}
\label{fig:result_all_extended2}
\end{figure*}

Figure~\ref{fig:result_all_extended} shows the stacked light profiles as a function of galactocentric radius (\textit{top} row) as well as surface brightness (\textit{bottom} row) for each of the sub-samples binned in galaxy stellar mass. For each case, in addition to our fiducial result, we also show the profiles computed assuming the SMC dust law (shown as \textit{dashed} curves) and the \citet{buat18} dust law (shown as \textit{dotted} curves). Figure~\ref{fig:result_all_extended2} shows the results for our two test dust-correction scenarios: \textit{(i)} assuming a single value of dust per galaxy (shown as \textit{dashed} curves); and \textit{(ii)} stacking only the pixels with a reliable per-pixel dust estimate (shown as \textit{dotted} curves). 

\begin{deluxetable}{c|D@{$\pm$}DD@{$\pm$}DD@{$\pm$}DD@{$\pm$}D}
\centering
\tablecaption{Best-fit slope parameters for trends in $\zeta$ presented in this work \label{tab:slope_values}.}
\tabletypesize{\small}
\tablehead{Sample & \multicolumn{4}{c}{$10^{8-11.5}$~M$_\odot$} & \multicolumn{4}{c}{$10^{8-9.5}$~M$_\odot$} & \multicolumn{4}{c}{$10^{9.5-10.2}$~M$_\odot$} & \multicolumn{4}{c}{$10^{10.2-11.5}$~M$_\odot$}}
\decimals
\startdata
 \hline
 \multicolumn{17}{c}{Slope as a function of galactocentric radius} \\
 \hline
 \textbf{$z$=0.7$-$1.5} &  0.111 & 0.054 & -0.014 & 0.087 &  0.022 & 0.083 &  0.224 & 0.104 \\
         $z$=0.7$-$1.1  &  0.198 & 0.047 &  0.028 & 0.072 &  0.122 & 0.086 &  0.273 & 0.068 \\
         $z$=0.1$-$1.5  & -0.019 & 0.092 & -0.052 & 0.154 & -0.056 & 0.115 &  0.049 & 0.145 \\
 \hline
 \multicolumn{17}{c}{Slope as a function of F140W surface brightness} \\
 \hline
 \textbf{$z$=0.7$-$1.5} &  0.054 & 0.019 & -0.073 & 0.040 & -0.032 & 0.028 &  0.126 & 0.030 \\
         $z$=0.7$-$1.1  &  0.117 & 0.017 & -0.095 & 0.019 &  0.006 & 0.026 &  0.177 & 0.025 \\
         $z$=0.1$-$1.5  &  0.012 & 0.043 & -0.056 & 0.082 & -0.056 & 0.057 &  0.085 & 0.084 \\
\enddata
\end{deluxetable}

\bibliographystyle{aasjournal}
\bibliography{Mehta_UVC_UVHa}

\end{document}